\documentclass[twocolumn, twocolappendix]{aastex701}

\newcommand\deltaB{\delta B/B_0}

\newcommand\skin{c/\omega_{\rm p}}
\newcommand\Lskin{L/(c/\omega_{\rm p})}
\newcommand\Kpar{K_\parallel}
\newcommand\Kparhat{\hat{K}_{\parallel, \, \ell}}
\newcommand\Kparhatnoell{\hat{K}_{\parallel}}
\newcommand\Kperp{K_\perp}
\newcommand\Ppar{P_\parallel}
\newcommand\Pperp{P_\perp}
\newcommand\betapar{\beta_\parallel}
\newcommand\rms{\rm rms}
\newcommand\Brms{B_{\rms}}
\newcommand\urms{u_{\rms}}
\newcommand\turnover{t_{\rm L}}
\newcommand\vth{v_{\rm th}}
\newcommand\gyro{r_{\rm g}}
\newcommand\gyrobar{\overline{r_{\rm g}}}
\newcommand\lambdacurve{\lambda_{\rm mfp, \, curve}}
\newcommand\contract{\omega_{\mathbf{K}}}

\usepackage{amsmath}
\usepackage{soul}

\received{\today}
\submitjournal{ApJ}

\begin{document}

\title{Intermittency in Collisionless Large-Amplitude Turbulence}

\author[orcid=0000-0001-6603-1983,gname='Ryan', sname='Golant']{Ryan Golant}
\affiliation{Department of Astronomy and Columbia Astrophysics Laboratory, Columbia University, New York, NY, 10027, USA}
\email[show]{ryan.golant@columbia.edu}  

\author[orcid=0000-0001-8822-8031,gname='Luca', sname='Comisso']{Luca Comisso}
\affiliation{Department of Astronomy and Columbia Astrophysics Laboratory, Columbia University, New York, NY, 10027, USA}
\email{lc3313@columbia.edu}  

\author[orcid=0009-0009-2144-3912,gname='Philipp', sname='Kempski']{Philipp Kempski}
\affiliation{Department of Astronomy and Columbia Astrophysics Laboratory, Columbia University, New York, NY, 10027, USA}
\email{pkempski@princeton.edu} 

\author[orcid=0000-0002-1227-2754,gname='Lorenzo', sname='Sironi']{Lorenzo Sironi}
\affiliation{Department of Astronomy and Columbia Astrophysics Laboratory, Columbia University, New York, NY, 10027, USA}
\affiliation{Center for Computational Astrophysics, Flatiron Institute, 162 5th Avenue, New York, NY 10010, USA}
\email{lsironi@astro.columbia.edu}

\begin{abstract}

Large-amplitude turbulence -- characterized by a fluctuating magnetic field component, $\delta B$, that is stronger than the mean component, $B_0$ -- is generically intermittent,  populated with intense localized structures such as sharp field-line bends and rapid field reversals. Recent MHD simulations suggest that these structures play an important role in particle transport and acceleration; however, MHD is inapplicable in most of our Universe, where the plasma is so hot or diffuse that Coulomb collisions are negligible. Therefore, in this paper, we analyze the intermittent properties of \emph{collisionless} large-amplitude turbulence in electron-positron plasmas via fully kinetic 3D simulations, exploring a wide range of $\deltaB$ and scale separations between the turbulence driving scale, $L$, and kinetic scales, $\skin$. The steady-state collisionless turbulence in our simulations broadly resembles that of MHD, but the development of pressure anisotropy steepens the scaling between magnetic field strength, $B$, and scalar field-line curvature, $\Kpar$ -- yielding $B \propto \Kpar^{-3/4}$ -- and consequently modifies the power-law slope of the probability density function of $\Kpar$; this slope hardens from $\Kpar^{-2.5}$ to $\Kpar^{-2.0}$ as $\deltaB$ increases from 4 to 140. Pressure anisotropy also triggers mirror and firehose instabilities, with the volume-filling fractions of these fluctuations increasing with $\deltaB$; for our largest $\deltaB$, $20\%$ of the volume is mirror-unstable and $6\%$ is firehose-unstable. Both the curvature and the Larmor-scale fluctuations in collisionless large-amplitude turbulence are expected to significantly influence cosmic ray transport and acceleration in the interstellar medium of our Galaxy and the intracluster medium of galaxy clusters.

\end{abstract}


\section{Introduction} \label{sec:introduction}
A wide array of astrophysical systems -- from planetary and stellar interiors to powerful accretion flows and winds to the interstellar and intracluster media -- exist in a state of magnetized turbulence \citep{biskamp_book, schek_book, davidson_book}. In this state, the chaotic advection, stretching, and folding of the magnetic field by turbulent motions is balanced by the back-reaction of the magnetic field via the Lorentz force. This nonlinear interplay between the velocity and magnetic fields dictates the morphology of the magnetic field and the shape of the magnetic power spectrum \citep{GoldreichSridhar1995, ChoLazarian2003, schek_dynamo, Boldyrev_2006, Beresnyak2019, Beattie2025}, the nature of cross-scale energy transfer and the details of small-scale dissipation \citep{dissipation_quataert}, and the efficacy of heat \citep{NarayanMedvedev2001, heating_dennis}, angular momentum \citep{momentum_shakura, BalbusHawley1998}, and particle transport \citep{transport_zweibel, feedback_zweibel, Lazarian2023}. As such, a comprehensive understanding of magnetized turbulence is a prerequisite to understanding much of our Universe. Unfortunately, a complete, unified theory of magnetized turbulence still eludes us \citep{schek_MHDreview}.     

One impediment to a complete theory of magnetized turbulence is our hazy understanding of intermittency, or the presence of sparse, intense, coherent structures -- such as current sheets, plasmoids, and sharp magnetic field-line bends -- arising from spatio-temporal nonuniformity in the turbulent energy cascade \citep{ intermittency_matthaeus, intermittency_vlahos, intermittency_ha}. Previous studies show that these intermittent structures can modulate how energy is dissipated \citep{Wan2012,Zhdankin2013, muni_tearing,Davis24}, how magnetic fields are amplified \citep{ galish_tearing, sironi_dynamo, muni_dynamo, beattie}, and how particles are scattered \citep{fielding, kempski_1, lemoine_transport} and accelerated \citep{Comisso18, Comisso19, Comisso22, joonas, lemoine_25}. Detrimentally, however, intermittency is absent from quasilinear models that treat turbulence as a sum of linear plane waves with uncorrelated random phases \citep{intermittency_marongoldreich}. Furthermore, because intermittent structures form prominently non-Gaussian distributions, they cannot be properly captured by two-point statistics like the power spectrum \citep{she_leveque}. Therefore, careful analysis of numerical simulations that go beyond simplistic synthetic wave turbulence is necessary to diagnose the impact of intermittency on magnetized turbulence. 

Recently, intermittency in the regime of large-amplitude magnetohydrodynamic (MHD) turbulence has received increased attention for its potential role in Galactic cosmic ray transport \citep{lemoine_transport, kempski_1, kempski_2, butsky, lubke}. This regime is characterized by a fluctuating magnetic field component, $\delta B$, that exceeds the mean component, $B_0$, due to the turbulent amplification of an initially weak 
field \citep{schek_dynamo}. In contrast to turbulence with $\deltaB \lesssim 1$, the magnetic field lines in large-amplitude turbulence are tangled and bent nearly isotropically (or fully isotropically in the $\delta B/B_0 \rightarrow \infty$ case, corresponding to the saturated state of the small-scale dynamo) \citep{schek_book}. The curvature of these field-line bends forms a wide power-law distribution cutting off near the resistive scale; this distribution has been hypothesized to make the transport of particles through the turbulence energy-dependent, since particles with smaller gyroradii can stay attached to field lines with tighter bends \citep{kempski_1, lemoine_transport}. MHD simulations show that the geometry of the magnetic field and the transport of cosmic rays in large-amplitude turbulence are sensitive to both $\delta B/B_0$ and the numerical grid resolution, but the detailed effects of these variables require further study \citep{kempski_2}. 

Crucially, however, the aforementioned picture of intermittency in large-amplitude MHD turbulence may break down in hot and low-density astrophysical environments where the mean free path to Coulomb collisions is large compared to the dynamical scales of interest; in such weakly collisional or collisionless environments, which include the intracluster medium and the hot interstellar medium, the collisional MHD description is no longer accurate and we must consider kinetic effects \citep{kunz_icm}. For instance, turbulent, magnetized, collisionless plasmas readily develop pressure anisotropy due to the conservation of particle magnetic moments (or, equivalently, conservation of the CGL invariants; \citet{cgl}): the pressure perpendicular to the magnetic field ($P_\perp$) exceeds the parallel pressure ($P_\parallel$) where the turbulent motions strengthen the field, while the parallel pressure dominates where the field is weakened.   

If the plasma $\beta$ -- the ratio of the thermal pressure to the magnetic pressure -- is high, then pressure anisotropy can trigger kinetic instabilities in collisionless plasmas \citep{arzamasskiy}. Such instabilities include the mirror \citep{Barnes66,Hasegawa69} and firehose \citep{Rosenbluth56, Chandrasekhar58, Parker58} instabilities, which grow preferentially on thermal Larmor scales and grow rapidly compared to fluid-scale flows, allowing these instabilities to significantly modify the turbulence \citep{instabilities_book, arzamasskiy}. In regions where $P_\perp/P_\parallel > 1 + 1/\beta$, the mirror instability causes field lines to bubble up, producing microscopic magnetic mirrors \citep{mirror}; in regions where $P_\perp/P_\parallel \lesssim 1 - 2/\beta$, the firehose instability causes Alfv\'en waves to grow exponentially, relaxing field-line tension \citep{firehose}. Not only do these instabilities alter the local geometry of the magnetic field, they can also suppress field-strength variations and anisotropic pressure stresses and introduce non-local energy transfer, ultimately changing the material properties of the plasma \citep{squire1, squire2, majeski}. Given these kinetic effects, collisionless large-amplitude turbulence may look much different from its MHD counterpart.       

Some initial insight into the general nature of collisionless large-amplitude turbulence can be gained by looking at previous simulations of the collisionless small-scale dynamo, the extreme ($\delta B/B_0 \rightarrow \infty$) limit of this regime \citep{rincon_smallscale, stonge_kunz, sironi_dynamo, muni_dynamo, radhika_dynamo}. In these simulations, mirror and firehose instabilities are triggered during the kinematic dynamo phase -- when the magnetic field is still too weak to affect the flow -- seeding small-scale magnetic fluctuations that provide an effective collisionality for the particles. Hence, even in the absence of Coulomb collisions, the plasma develops fluid-like behavior. In saturation, the magnetic field forms a large-scale folded sheet structure reminiscent of the equivalent MHD dynamo \citep{schek_dynamo}; on small scales, however, the magnetic field is locally corrugated by nonlinear mirrors and firehoses. Whether this same general agreement with MHD holds at smaller $\delta B/B_0 > 1$ is a central focus of this work.  

In this paper, we analyze the intermittent properties of moderately subsonic, large-amplitude turbulence in a collisionless electron-positron plasma via an extensive suite of fully kinetic 3D simulations covering a wide range of $\deltaB$ (from $\deltaB \approx 1$ up to $\deltaB \approx 140$) and of scale separation between the turbulent driving scale and kinetic scales (using simulation boxes with $L/\left(\skin\right) = \left\{500, 1000, 2000\right\}$, for $\skin$ the plasma skin depth and $L$ the box size, which is also our driving scale). We primarily focus on the statistics of magnetic field-line curvature -- which has direct connections to particle transport and acceleration (e.g., \citet{kempski_1, kempski_2, lemoine_25}) -- probing how kinetic effects yield deviations from MHD. Broadly, the steady-state turbulence in our collisionless simulations resembles that of MHD, as had previously been seen in the case of the collisionless small-scale dynamo. However, the development of pressure anisotropy steepens the scaling between magnetic field strength and scalar field-line curvature and consequently modifies the power-law slope of the probability density function of curvature. Pressure anisotropy also triggers mirror and firehose instabilities in each of our simulations, with the volume-filling fractions of these fluctuations increasing with $\deltaB$. Both the curvature and the Larmor-scale fluctuations in collisionless large-amplitude turbulence are expected to significantly influence cosmic ray transport and acceleration in $\beta \gtrsim1$ plasmas such as the hot interstellar medium of our Galaxy and the intracluster medium of galaxy clusters. 

This paper is structured as follows: in Section \ref{sec:setup} we detail the setup of our simulation suite; in Section \ref{sec:properties} we provide an overview of the salient properties of the steady-state turbulence in our simulations; in Section \ref{sec:statistics} we analyze the statistics of magnetic field-line curvature; in Section \ref{sec:anisotropy} we show how pressure anisotropy modifies the curvature statistics; in Section \ref{sec:discussion} we connect our results to particle transport and acceleration; and in Section \ref{sec:conclusions} we summarize our main results and conclude.

\section{Simulation Setup} \label{sec:setup}

We run fully periodic 3D simulations of moderately subsonic, large-amplitude turbulence in a collisionless electron-positron plasma using the electromagnetic particle-in-cell (PIC) code TRISTAN-MP \citep{tristan_buneman, tristan}. The code solves the Maxwell-Vlasov system using a second-order Finite-Difference Time-Domain (FDTD) scheme on a standard Yee mesh \citep{yee} and pushes particles using the Boris algorithm \citep{boris} and first-order shape functions. The electromagnetic fields are extrapolated to the particle positions using a trilinear interpolation function. At each time step, after depositing the electric current to the grid using the charge-conserving zig-zag scheme \citep{zigzag}, we apply ten passes of a 3-point (1-2-1) digital current filter in each direction to smooth out non-physical short-wavelength oscillations. The simulations analyzed in the main text use 8 particles per cell per species and a grid resolution of 1 plasma skin depth, $\skin \equiv \left(m_{\rm e} c^2/4\pi n_0 e^2\right)^{1/2}$, where $n_0$ is the uniform initial density of the plasma (including both species) and $c$ is the speed of light, set to 0.45 cells per timestep.

To drive turbulence, we add a charge-independent external acceleration term, $\mathbf{a}(\mathbf{x},t)$, to the particle equations of motion \citep{sironi_dynamo}. This acceleration field is decomposed into six independently evolving Fourier modes such that 
\begin{equation} \label{eq:acceleration}
    \mathbf{a}(\mathbf{x},t) = \Re\left[\sum_\mathbf{k} \mathbf{\tilde{a}}(\mathbf{k},t) e^{i\mathbf{k} \cdot \mathbf{x}} \right],
\end{equation}
where the sum is over all six modes, each driven at the simulation box scale, $L$: $\mathbf{k}L/2\pi = \left\{(\pm 1,0,0), (0, \pm 1, 0), (0,0,\pm 1) \right\}$. $\mathbf{\tilde{a}}(\mathbf{k},t)$ in Equation \eqref{eq:acceleration} is the amplitude of mode $\mathbf{k}$, advanced in Fourier space according to an Ornstein-Uhlenbeck process \citep{uhlenbeck_ornstein_1930,zrake_driving}:
\begin{equation} \label{eq:ornstein}
    d\mathbf{\tilde{a}}(\mathbf{k},t) = -\gamma_{\rm corr} \mathbf{\tilde{a}}(\mathbf{k},t) dt + \sqrt{\gamma_{\rm corr} P_k} \mathbf{T}(\mathbf{k}) \cdot d\mathbf{\tilde{W}}(\mathbf{k},t).
\end{equation}
The first term on the right-hand side of this equation functions as a ``drag'' term pulling $\mathbf{\tilde{a}}(\mathbf{k},t)$ towards its mean, with the efficacy of this drag set by the decorrelation rate $\gamma_{\rm corr} \simeq 1.5/\turnover$ (for $\turnover$ the turnover time); this drag modulates the Wiener process (i.e., the continuous random walk) modeled by the second term on the right-hand side. In this second term, $d\mathbf{\tilde{W}}(\mathbf{k},t)$ is a complex random variable with real and imaginary parts each uniformly distributed within $\left[-1/2, 1/2\right]$, while $\mathbf{T}(\mathbf{k})$ is a projection operator that dictates the fraction of each driven mode that goes into compressive vs. shearing/solenoidal motions,
\begin{equation} \label{eq:project}
    \mathbf{T}(\mathbf{k}) \equiv (1-\zeta)\mathbf{T_\parallel}(\mathbf{k}) + \zeta\mathbf{T_\perp}(\mathbf{k}),
\end{equation}
where 
\begin{equation}
    \mathbf{T}_\parallel(\mathbf{k}) \cdot d\mathbf{\tilde{W}}(\mathbf{k},t) = \left[\mathbf{\hat{k}} \cdot \mathbf{\tilde{W}}(\mathbf{k},t) \right] \mathbf{\hat{k}}
\end{equation}
and 
\begin{equation}
    \mathbf{T}_\perp(\mathbf{k}) \cdot d\mathbf{\tilde{W}}(\mathbf{k},t) = d\mathbf{\tilde{W}} - \left[\mathbf{\hat{k}} \cdot \mathbf{\tilde{W}}(\mathbf{k},t) \right].
\end{equation}
In Equation \ref{eq:project}, $\zeta = 0$ corresponds to purely compressive driving and $\zeta = 1$ corresponds to purely solenoidal driving. All of the simulations analyzed in this paper use purely solenoidal driving. \footnote{While compressive modes may be important for particle acceleration (e.g., \cite{zhdankin_accel, BrunettiLazarian2007}), previous simulations of MHD turbulence suggest that these modes have a subdominant impact on particle transport, accounting for only a small fraction of the turbulent kinetic energy \citep{kempski_1, Gan2022}.} The factor $P_k$ in Equation \eqref{eq:ornstein}, which is the same for all modes, serves as a normalization accounting for the different degrees of freedom in compressive vs. solenoidal modes; more precisely, $P_k \propto \left[(1-\zeta)^2 + 2\zeta^2\right]^{-1}$, where the constant of proportionality sets the target root-mean-squared velocity of the turbulent driver, $u_{{\rm rms},0}$.

We choose numerical parameters such that the turbulent driver produces a non-relativistic and moderately subsonic flow. The plasma is initialized from a Maxwellian distribution with $k_{\rm B} T_0/m_{\rm e} c^2 = 0.04$, where $T_0$ is the initial temperature of both the electrons and the positrons; this implies a thermal speed of $v_{\rm th} \equiv \sqrt{2k_B T_0/m_e} \approx 0.3 c$. The plasma temperature remains roughly constant due to our particle cooling prescription: to keep the system in quasi-steady state, we artificially cool particles that exceed a threshold momentum of $p_{\rm cut}/m_{\rm e} c = 0.7$, resetting the momentum to $p_{\rm cut}$ and keeping the particle's direction unchanged. By setting the target rms velocity of our driver to $u_{\rm rms,0} = 0.2 c$, the sonic Mach number of the turbulence oscillates around a steady value, $M_{\rm s} \equiv u_{\rm rms}/v_{\rm th} \approx 0.7$ (or $u_{\rm rms}/c_s \approx 0.8$ for $c_s \equiv \sqrt{(5/3)k_B T_0/m_e}$), where $\urms$ is the instantaneous rms velocity of the flow; therefore, our turbulence remains moderately subsonic for the majority of its time-evolution (see Figure \ref{fig:energies_2e3}). By normalizing the Mach number to $\vth$ (which is standard in the collisionless dynamo literature, e.g., \citet{rincon_smallscale, stonge_kunz, muni_dynamo, radhika_dynamo}), $M^2$ yields a measurement of the ratio of turbulent kinetic energy to thermal energy:
\begin{equation} \label{eq:mach2}
    M_{\rm s}^2 = \frac{\frac{1}{2} m_e \urms^2 }{k_{\rm B} T_0}.
\end{equation}

The key physical parameters that we vary throughout our simulation suite are the initial mean magnetic field strength, $B_0$, and the scale separation between the driving scale and kinetic scales, $L/(\skin)$, set by the simulation box size, $L$. We initialize each simulation with a seed magnetic field pointing in the $z$ direction with strength set by the initial inverse plasma $\beta$,
\begin{equation} \label{eq:beta0}
    \beta_0^{-1} \equiv \frac{B_0^2}{8\pi n_0 k_{\rm B} T_0}.
\end{equation}
We evolve our simulations until the magnetic field amplification saturates and the turbulence reaches a steady state, as illustrated in Figure \ref{fig:energies_2e3}. In steady state, we quantify the strength of the turbulence via the ratio of the rms strength of the fluctuating component of the magnetic field to the strength of the mean field,
\begin{equation} \label{eq:deltaB}
    \frac{\delta B}{B_0} \equiv \frac{\sqrt{\left<\delta B^2 \right>}}{B_0},
\end{equation} 
where angled brackets indicate averaging over the whole volume. Throughout this paper, we use $\deltaB$ to label our simulations; in Table \ref{tab:simulations}, we show the correspondence between our input mean-field strengths (i.e, $\beta_0^{-1}$), the measured (time-averaged) $\deltaB$ in steady-state, and the measured (time- and spatial-averaged) plasma $\beta$ in steady state, computed as
\begin{equation} \label{eq:beta}
    \beta \equiv \frac{8\pi n_0 k_{\rm B} T_0}{\left<B^2\right>},
\end{equation}
for $B$ the local magnetic field strength. Coupled with Equation \eqref{eq:mach2}, $M_{\rm s}^2 \, \beta \sim 1$ provides a convenient condition for energy equipartition.

\begin{table}
    \centering
    \begin{tabular}{|c|c|c|}
        \hline
        $\beta_0^{-1}$ & $\deltaB$ & $ \beta$ \\
        \hline
         $3.75 \times 10^{-1}$ & 1 & 1.2\\
         $1.25 \times 10^{-1}$ & 2 & 1.6\\
         $3.75 \times 10^{-2}$ & 4 & 1.8 \\
         $1.25 \times 10^{-2}$ & 6 & 1.8 \\
         $3.75 \times 10^{-3}$ & 10 & 2.4 \\
         $1.25 \times 10^{-3}$ & 16 & 3.0\\
         $1.25 \times 10^{-4}$ & 45 & 3.6\\
         $1.25 \times 10^{-5}$ & 140 & 3.5\\
         0 & $\infty$ & 3.8 \\
         \hline
    \end{tabular}
    \caption{Correspondence between the initial mean-field strength, $\beta_0^{-1}$ (Equation \eqref{eq:beta0}), the measured amplitude of the resulting steady-state turbulence, $\deltaB$ (Equation \eqref{eq:deltaB}), and the measured steady-state plasma $\beta$ (Equation \eqref{eq:beta}) for each of our simulations. The values of both $\deltaB$ and $\beta$ have been time-averaged across the steady-state duration.}
    \label{tab:simulations}
\end{table}

For each $\deltaB$, we run three simulations with boxes of size $L^3 = \left(500 \, \skin\right)^3$, $\left(1000 \, \skin\right)^3$, and $\left(2000 \, \skin\right)^3$ to check the dependence of the turbulence on scale separation between the driving scale and kinetic scales, $L/\left(\skin\right)$. Unless otherwise noted, the data shown in the main text of this paper correspond to our largest (i.e., $\left(2000 \, \skin\right)^3$) boxes.

To smooth out statistical fluctuations in our measurements of the turbulence, we average our results over at least five snapshots in steady state, each spaced $9 \, \turnover$ apart; all of the figures in this paper (except for the time series and simulation-box visualizations) show time-averaged quantities. Additionally, to ensure that our measurements of intermittency are not affected by small-scale noise -- which poses a particular problem when numerically computing spatial gradients -- we filter out the magnetic field at scales below $\sim 9 \, \skin$, corresponding to the scale at which the magnetic power spectrum becomes noise-dominated (see Section \ref{sec:properties}); to do this, we Fourier transform the magnetic field, set all modes with wavenumber greater than $0.7 \, \left(\skin\right)^{-1}$ to $0$, then inverse transform back to real space. This filtering makes our measurements of the magnetic field-line geometry more robust.

\section{General properties of the turbulence in steady state} \label{sec:properties}

Throughout this paper, we discuss the properties of our simulated turbulence in steady state. To start, therefore, it is instructive to describe the evolution of our simulations towards this steady state.

\begin{figure}[h]
    \includegraphics[width=0.5\textwidth]{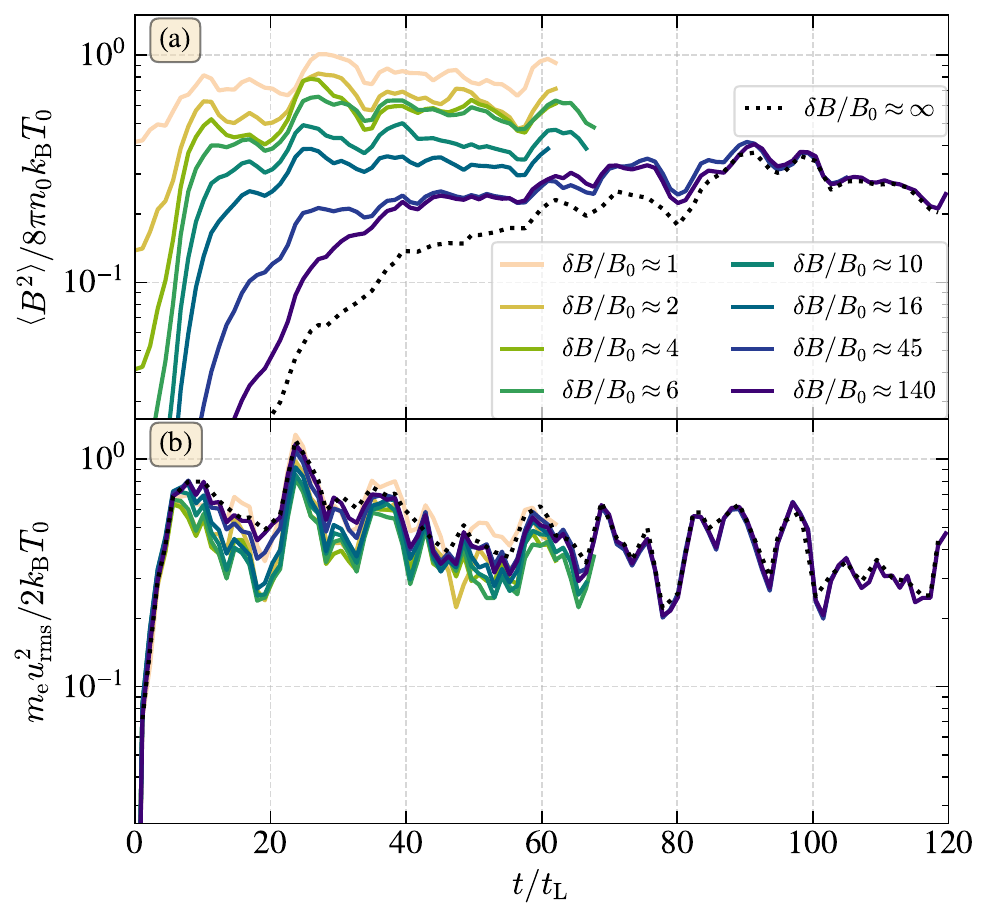}
    \caption{Time evolution of the magnetic energy (top panel) and the kinetic energy (bottom panel) for each of our simulations with $\Lskin = 2000$; the unmagnetized dynamo case, $\deltaB \rightarrow \infty$, is depicted with a dotted curve to distinguish it from the rest of the cases. The magnetic energy is normalized to give $1/\beta$ at fixed temperature (Equation \eqref{eq:beta}) and the kinetic energy is normalized to give the square of the sonic Mach number at fixed temperature (Equation \eqref{eq:mach2}). Each simulation reaches a steady state within a few tens of eddy turnover times, $\turnover \equiv L/2\pi u_{\rm rms,0}$.}
    \label{fig:energies_2e3}
\end{figure}

Figure \ref{fig:energies_2e3} shows the full time-evolution of the magnetic energy (top panel) and of the kinetic energy (bottom panel) for each of our simulations with $\Lskin = 2000$; darker colors correspond to larger $\deltaB$. For every $\deltaB$, the initial magnetic field experiences some degree of amplification that saturates within a few tens of $\turnover \equiv L/2\pi u_{\rm rms,0}$ (with the time of saturation dependent on $\beta^{-1}_0$), yielding near-equipartition between the magnetic and kinetic energies. This is the first time that turbulent magnetic field amplification has been demonstrated in fully-kinetic collisionless turbulence with a non-zero mean field; while previous work showed that the early growth of the magnetic field in the $B_0 = 0$ (i.e., $\deltaB \rightarrow \infty$) case was due purely to fields seeded by the Weibel instability \citep{sironi_dynamo, muni_dynamo}, it remains the subject of future work to describe the role of collisionless effects at early times for larger $B_0$ (i.e., smaller $\deltaB$). This early physics may explain the bifurcation we see in magnetic field evolution between $\deltaB \approx 16$ and $\deltaB \approx 45$: the larger $\deltaB$ cases ($\deltaB \gtrsim 45$) grow gradually, clustering around the $\deltaB \rightarrow \infty$ case by $90 \, \turnover$, while the smaller $\deltaB$ cases ($\deltaB \lesssim 16$) grow rapidly, reaching saturation by $20 \, \turnover$. These two regimes are evident across a diverse array of measurements, as we shall see throughout this paper. Because the steady-state properties of the larger $\deltaB$ simulations all closely resemble those of the $\deltaB \rightarrow \infty$ simulation, we discard the $\deltaB \rightarrow \infty$ case from further plots; see \citet{sironi_dynamo} for further analysis of this case. 

The magnetic energy in Figure \ref{fig:energies_2e3} is normalized to give $1/\beta$, for $\beta$ defined in Equation \eqref{eq:beta}. As illustrated in this plot and in Table \ref{tab:simulations}, our simulations span a wide range of $\beta$ in steady state: the smaller $\deltaB$ cases reach $\beta$ of order unity, characteristic of the interstellar medium of galaxies, while the larger $\deltaB$ cases develop higher-$\beta$ turbulence, more characteristic of the intracluster medium of galaxy clusters.

In the bottom panel of Figure \ref{fig:energies_2e3}, the kinetic energy is normalized to give the square of the sonic Mach number, $M_{\rm s}^2$, under the assumption of constant temperature, as defined in Equation \eqref{eq:mach2}. Evidently, the strength of the mean magnetic field has only a minor impact on the development of the turbulent flow: for every $\deltaB$, the rms velocity quickly reaches a steady state (within $10 \, \turnover$) that fluctuates around the target $u_{\rm rms, 0}$ of $0.2 c$, or around $M_s \approx 0.7$. Therefore, the turbulence in each of our simulations remains moderately subsonic throughout the majority of its time-evolution. The largest Alfv\'en speed reached in our simulations is $v_{\rm A} \approx 0.3 c$, making our turbulence safely non-relativistic.

\begin{figure}[h]
    \includegraphics[width=0.5\textwidth]{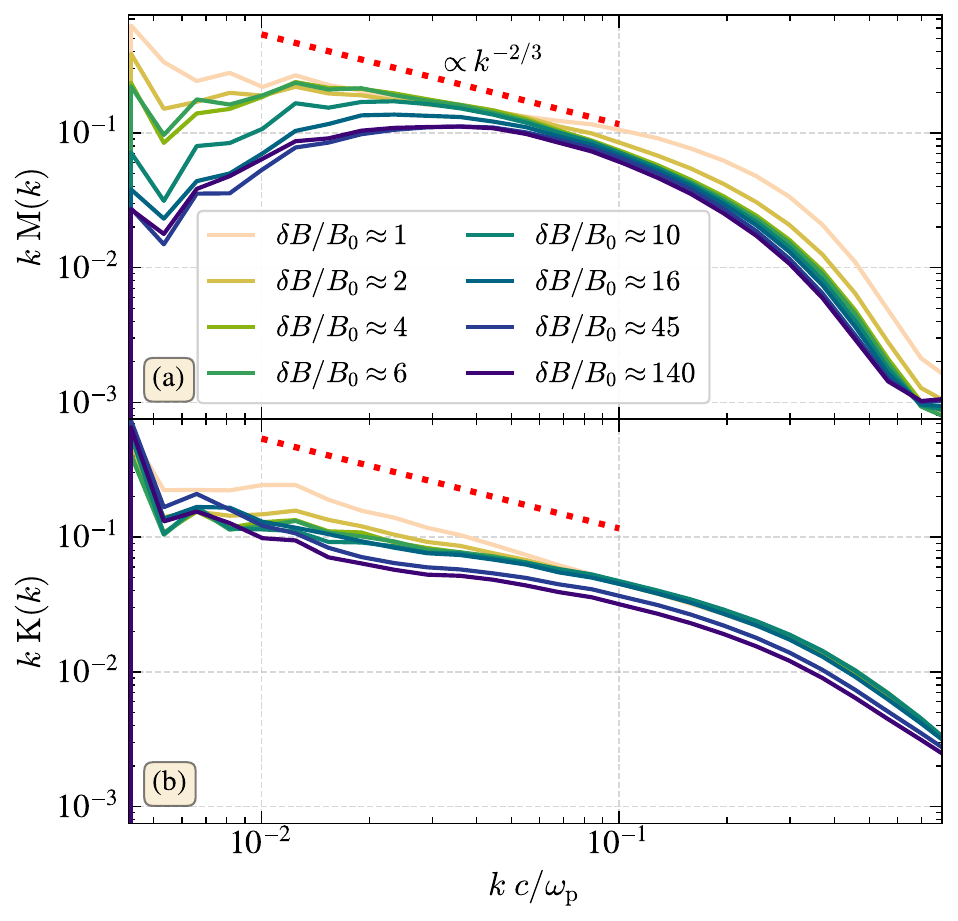}
    \caption{Isotropic magnetic (top panel) and kinetic (bottom panel) power spectra for each of our simulations with $\Lskin = 2000$. The normalization of each spectrum is multiplied by the radial wavenumber, $k$, to show the energy in log-spaced bins; the integral over the magnetic spectrum yields $1/\beta$ (Equation \eqref{eq:beta}) and the integral over the kinetic spectrum yields the square of the sonic Mach number (Equation \eqref{eq:mach2}). The red dotted lines in each panel indicate the Kolmogorov scaling ${\rm M}(k) \propto {\rm K}(k) \propto k^{-5/3}$ \citep{kolmogorov}. The magnetic power spectrum becomes noise-dominated at wavenumbers $k \, \skin \gtrsim 0.7$, or physical scales below $\sim 9 \, \skin$.}
    \label{fig:spectra_2e3}
\end{figure}

To understand the distribution of energy as a function of spatial scale in our developed turbulence, we plot the steady-state isotropic magnetic (top panel) and kinetic (bottom panel) power spectra, ${\rm M}(k)$ and ${\rm K}(k)$, respectively, in Figure \ref{fig:spectra_2e3}; the spectra are normalized such that the integral over the magnetic spectrum yields $1/\beta$ (Equation \eqref{eq:beta}) and the integral over the kinetic spectrum yields $M_{\rm s}^2$ (Equation \eqref{eq:mach2}). The bifurcation between large $\deltaB$ and small $\deltaB$ is also evident in these spectra: the large $\deltaB$ cases produce dynamo-like spectra with most of the magnetic energy residing well below the outer scale (peaking around $k \, \skin \sim 0.04$), while the small $\deltaB$ cases have pronounced peaks at the outer scale (due largely to the mean field), forming Kolmogorov-like spectra with slopes close to $-5/3$ persisting across a well-defined inertial range ($10^{-2} \lesssim k \, \skin \lesssim 10^{-1}$) \citep{kolmogorov}. 

We can quantify the dominant scale of the magnetic field via the integral-scale wavenumber,
\begin{equation} \label{eq:kint}
    k_{\rm int} \equiv \frac{\int {\rm M}(k) \, {\rm d}k}{\int k^{-1} \, {\rm M}(k) \, {\rm d}k};
\end{equation}
while the $\deltaB \gtrsim 16$ cases each have $k_{\rm int} \, \skin \sim 2 \times 10^{-2}$ (or a physical scale, $L_{\rm int} \equiv 2\pi/k_{\rm int}$, around one-sixth of the box scale, $L$), $k_{\rm int}$ steadily decreases with lower $\deltaB$, reaching $k_{\rm int} \, \skin \sim 10^{-2}$ (or $L_{\rm int} \sim L/3$) for $\deltaB \approx 1$. $L_{\rm int}$ increases linearly with box size: as shown in Figure \ref{fig:kints}, increasing $\Lskin$ by a factor of two decreases $k_{\rm int}$ (and increases $L_{\rm int}$) also by a factor of two.   

As illustrated in the bottom panel of Figure \ref{fig:spectra_2e3}, each of our simulations develops a full turbulent cascade running from the box scale down to the grid scale. The steep drop in the kinetic spectra at low $k$ ($k \, \skin \sim 0.004-0.005$) indicates that not all kinetic energy injected at the driving scale cascades to smaller scales; indeed, for all $\deltaB$, the kinetic energy remains subdominant to the magnetic energy throughout the inertial range ($10^{-2} \lesssim k \, \skin \lesssim 10^{-1}$). However, the kinetic spectra fall off more gradually than the magnetic spectra at high $k$ ($k \, \skin \gtrsim 10^{-1}$); the kinetic spectra steepen mildly as they enter a dissipative range around $k \, \skin \sim 3 \times 10^{-1}$, while the magnetic spectra steepen dramatically beyond $k \, \skin \sim 2 \times 10^{-1}$, with this magnetic dissipative cutoff shifting towards higher $k$ for larger $\deltaB$. 

% what does magnetic spectrum look like without mean field? does this explain dissipative scale of low deltaB/B cases (i.e., why high-k stuff in spectra?). how does contribution from mean field compare to contribution from dynamo amplification? how does this affect peak of spectrum (at some point mean field dominates). Something about tearing mediated dynamo and presence of plasmoids/current sheets?

% tearing and prandtl dependence?

% kinetic spectra: Not all energy cascades from large to small scales (dip in kinetic spectrum);

\begin{figure*}[h]
    \includegraphics[width=1.0\textwidth]{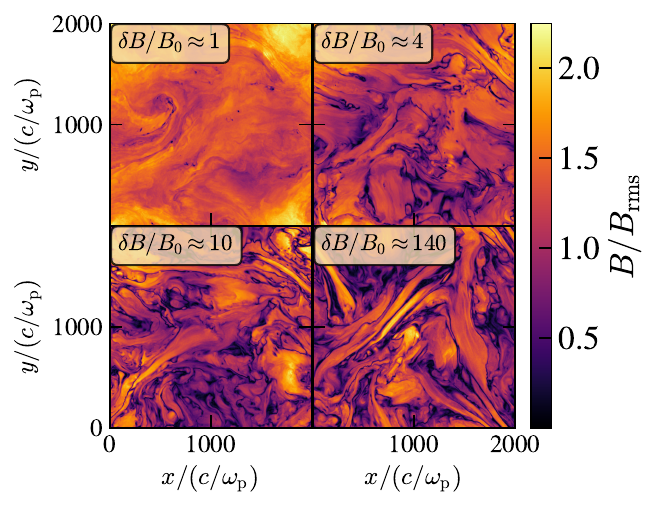}
    \caption{2D slices (in the $xy$-plane) of $B/\Brms$ -- the local magnetic field strength normalized by the global rms magnetic field strength -- at late times for our simulations with $\deltaB \approx  1$ (upper left), $4$ (upper right), $10$ (lower left), and $140$ (lower right) and $\Lskin = 2000$; brighter colors represent regions of stronger field.}
    \label{fig:volume_2e3}
\end{figure*}

The multi-scale nature of the steady-state turbulent magnetic field, quantified in the spectra of Figure \ref{fig:spectra_2e3}, can be seen qualitatively in Figure \ref{fig:volume_2e3}, which shows 2D slices (in the $xy$-plane) of $B/\Brms$ -- the local magnetic field strength normalized by the global rms magnetic field strength, $\Brms \equiv \sqrt{\left<B^2 \right>}$ -- in our $\deltaB \approx 1$ (upper left), $4$ (upper right), $10$ (lower left), and $140$ (bottom right) simulation boxes at late times. For $\deltaB \approx 140$, the flow has arranged the magnetic field into a folded sheet structure reminiscent of the saturated MHD dynamo \citep{schek_dynamo}; as $\deltaB$ decreases, the length of these structures decreases, yielding a more well-mixed turbulence. These turbulent magnetic fields also possess abundant substructure at smaller scales. For example, the folding of field-lines has created an entire distribution of bent magnetic structures, with the smallest radii of curvature comparable to the magnetic dissipative scale. Additionally, current sheets (across which the magnetic field reverses rapidly) and reconnection plasmoids are especially prominent at larger $\deltaB$.   
% discuss volume renderings. this turbulence corresponds to the steady states shown in energy evolution. This also allows us to visualize multi-scale phenomena reflected in magnetic spectrum. Folded fields corrugated by small-scale features. More turbulent for higher sigma?

% \begin{figure}[h]
%     \includegraphics[width=0.5\textwidth]{fig_volumeB.png}
%     \caption{[this is not the final figure] Surface plots of the magnetic field strength, $B/\Brms$, at late times for our simulations with $\deltaB = 1$, $10$, and $140$ and $\Lskin = 2000$; brighter colors represent regions of stronger field. A few prominent magnetic field reversals are indicated with white arrows, while a few reconnection plasmoids are indicated with green arrows.}
%     \label{fig:volume_2e3}
% \end{figure}

\begin{figure*}
    \includegraphics[width=1\textwidth]{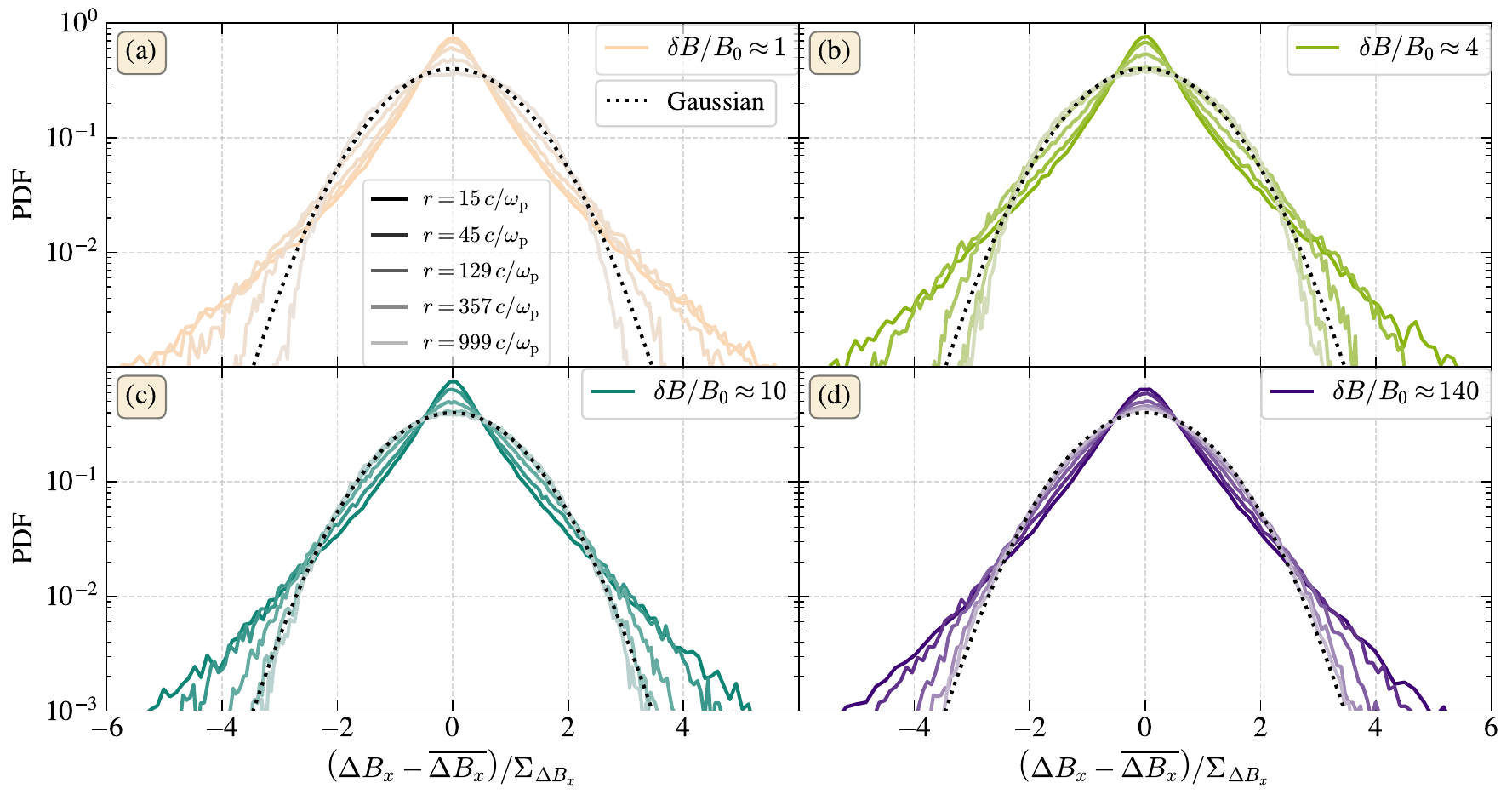}
    \caption{Renormalized probability density functions (PDFs) of increments of the $x$-component of the magnetic field (Equation \eqref{eq:increment}) for our simulations with $\deltaB = 1$ (upper left), $4$ (upper right), $10$ (lower left), and $140$ (lower right) and $\Lskin=2000$. By subtracting the mean of the raw PDF, $\overline{\Delta B_x}$, and dividing by the standard deviation of the raw PDF, $\Sigma_{\Delta B_x}$, these renormalized PDFs have a mean of zero and a standard deviation of unity; the equivalent Gaussian distribution is shown as a dotted curve in each panel. For each $\deltaB$, we show the renormalized PDF of field increments at five different separations, $r/(\skin) = 15$, $45$, $129$, $357$, and $999$, with lighter shades for larger separations.}
    \label{fig:deltabhist_2e3}
\end{figure*}

The coherent structures at small scales in Figure \ref{fig:volume_2e3} hint at the presence of intermittency in the turbulent magnetic field, or deviation of $B$ from a Gaussian distribution. We quantify this in Figure \ref{fig:deltabhist_2e3}, which shows renormalized probability density functions (PDFs) of increments of the $x$-component of the magnetic field \footnote{Our decision to take increments of $B_x$ is arbitrary -- we reach the same conclusions by taking increments of $B_y$ or $B_z$.},
\begin{equation} \label{eq:increment}
    \Delta B_x(r) \equiv \left|B_x(\mathbf{x}+\mathbf{r}) - B_x(\mathbf{x})\right|,
\end{equation}
for $\deltaB \approx$ 1 (upper left), 4 (upper right), 10 (lower left), and 140 (lower right); for each $\deltaB$, we sample increments at separations of $r/(\skin) = 15$, $45$, $129$, $357$, and $999$ (with lighter shades for larger $r$) to probe structure at a wide range of scales. We also subtract off the mean of each of our PDFs, $\overline{\Delta B_x}$, and divide by the standard deviation, $\Sigma_{\Delta B_x}$, so that our final distributions have a mean of zero and standard deviation of unity. As expected for an intermittent field, our renormalized PDFs increasingly deviate from Gaussian as we go to smaller scales (i.e., smaller separations, $r$) \citep{she_leveque}; at large scales the field is still sensitive to the driver (a Gaussian process), but nonuniformity in the turbulent cascade intensifies towards smaller scales, producing progressively more intense localized structures.
% Comparing the four $\deltaB$ cases in Figure \ref{fig:deltabhist_2e3}, it appears that the magnetic field becomes more intermittent with smaller $\deltaB$, with even the large scales showing non-Gaussianity in the $\deltaB \approx 1$ and $\deltaB \approx 4$ cases. 
The widespread presence of non-Gaussianity in our simulations motivates our study of intermittency in collisionless large-amplitude turbulence.

\section{Statistics of magnetic field-line bends and reversals} \label{sec:statistics}

In the previous section, we showed that the magnetic field in collisionless large-amplitude turbulence is generically intermittent. Here, we focus on a specific class of intermittent structures present in the magnetic field: sharp magnetic field-line bends and rapid field reversals. These structures are formed by the stretching and folding of field lines into elongated geometries, across which the sign of the magnetic field varies over short length scales; at the ends of these structures, the magnetic field lines are tightly curved. These structures have direct implications for particle transport and particle acceleration, which we explore in more depth in Section \ref{sec:discussion}.

We can characterize field-line bends and reversals with two quantities: the magnitude of the field-line curvature, $\Kpar$, and the inverse reversal scale, $\Kperp$ \citep{kempski_1}. Typically, the field-line curvature is defined as $\left|\mathbf{\hat{b}} \cdot \mathbf{\nabla\hat{b}}\right|$, where $\mathbf{\hat{b}} \equiv \mathbf{B}/B$ is the magnetic field unit vector; this can be understood as the variation of the magnetic field direction along the magnetic field itself. However, in this paper, we compute a slightly different quantity, removing any numerically introduced components of curvature along the field \citep{kempski_1}:
\begin{equation} \label{eq:kpar}
    \Kpar \equiv \left|\mathbf{\hat{b}} \times 
    \left(\mathbf{\hat{b}} \cdot \mathbf{\nabla\hat{b}}\right)\right|.
\end{equation}
We do not find significant discrepancy in our results when using $\Kpar$ vs. the traditional definition of curvature, but we do find that $\Kpar$ is more robust to numerical artifacts like grid downsampling.

To complement $\Kpar$, which quantifies variation along the field, we define $\Kperp$ to quantify variation across (or perpendicular to) the field. A common metric used to quantify variation across the field in the dynamo literature (e.g., \cite{schek_dynamo, galish_tearing}) is 
\begin{equation}
    k_{\mathbf{J} \times \mathbf{B}} \equiv \frac{\left|\mathbf{J} \times \mathbf{B}\right|}{B^2} = \frac{\left|\left(\mathbf{\nabla} \times \mathbf{B}\right) \times \mathbf{B}\right|}{B^2}.
\end{equation} 
However, we follow \citet{kempski_1} by defining
\begin{equation} \label{eq:kperp}
    \Kperp \equiv k_{\mathbf{J} \times \mathbf{B}} - \Kpar = \left|\mathbf{\hat{b}} \times \left(\mathbf{\hat{b}} \times \nabla \ln B  \right)\right|.
\end{equation}
At a field-line bend, $\Kpar \sim \Kperp$.

\begin{figure*}
    \includegraphics[width=1\textwidth]{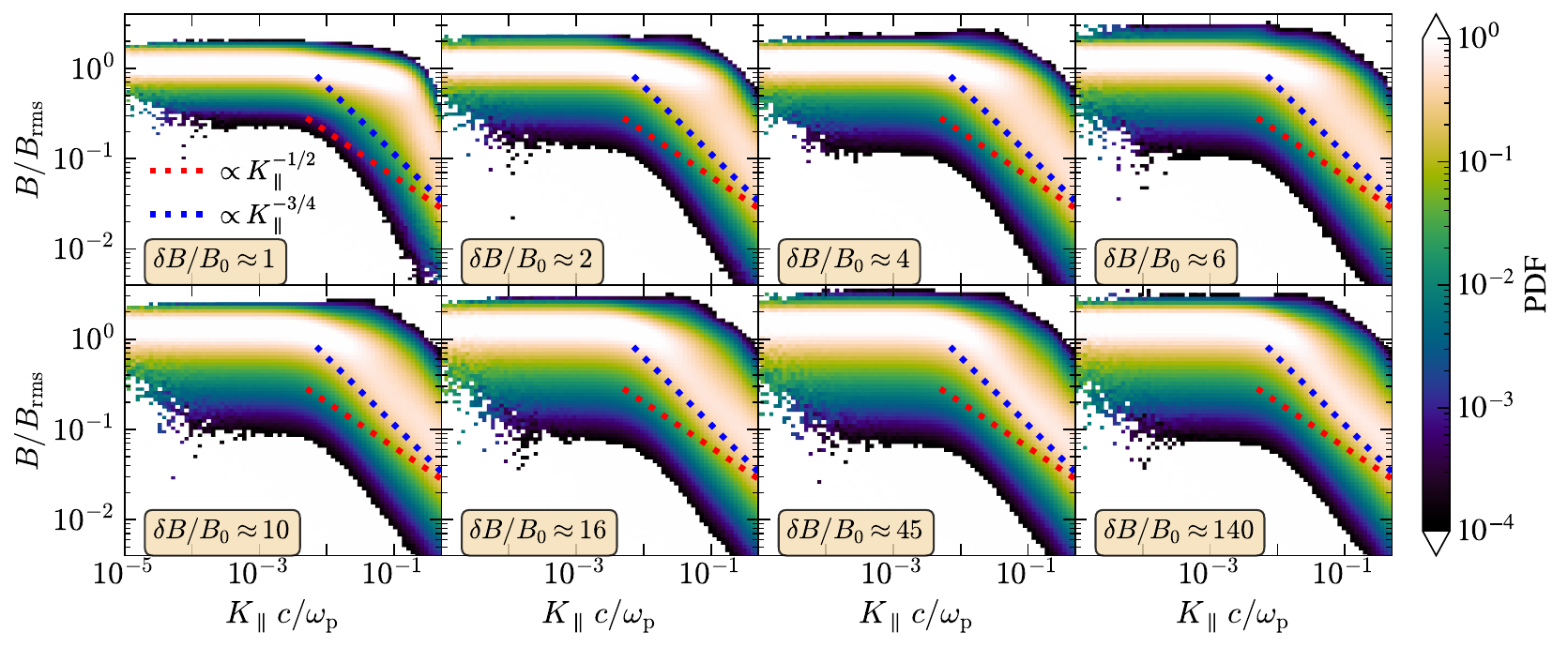}
    \caption{2D PDFs of the local magnetic field strength, $B/\Brms$, vs. the local field-line curvature, $\Kpar$, for each of our simulations with $\Lskin=2000$ (see panel labels). The red dotted line in each panel shows the scaling $B \propto \Kpar^{-1/2}$ (expected in MHD, e.g., \citet{schek_dynamo}), while the blue dotted line shows $B \propto \Kpar^{-3/4}$. To increase the contrast in the high-$\Kpar$ tails of these PDFs, we normalize the PDFs such that the integral over the 1D PDF of $B/\Brms$ at each $\Kpar$ is unity.}
    \label{fig:bcurve_2e3}
\end{figure*}

\begin{figure}[h]
    \includegraphics[width=0.5\textwidth]{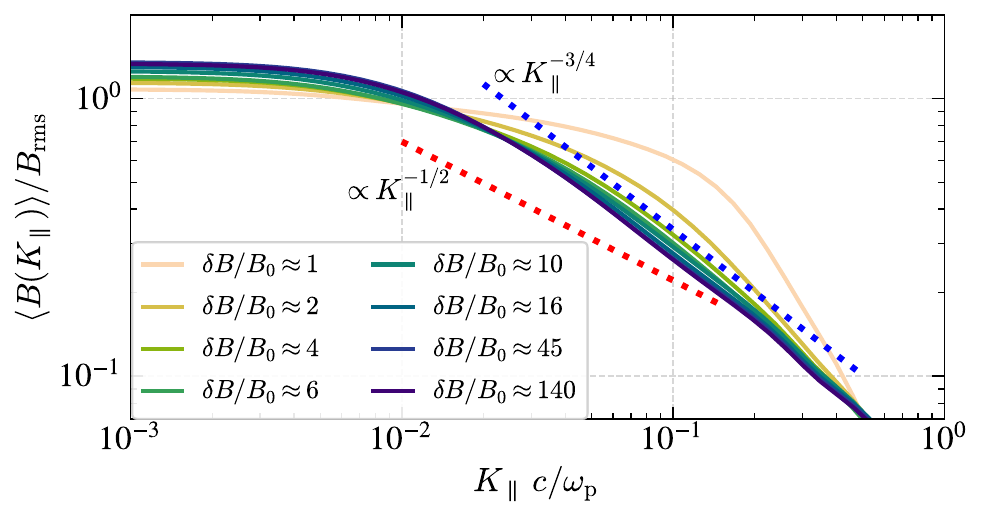}
    \caption{Average magnetic field strength, $\left<B(\Kpar)\right>/\Brms$, as a function of field-line curvature, $\Kpar$, for each of our simulations with $\Lskin=2000$. The red dotted line shows the scaling $B \propto \Kpar^{-1/2}$, while the blue dotted line shows $B \propto \Kpar^{-3/4}$.}
    \label{fig:bcurve1D_2e3}
\end{figure}

To begin our analysis of field-line curvature in collisionless large-amplitude turbulence, we show 2D PDFs of local magnetic field strength ($B/\Brms$) vs. $\Kpar$ for each of our simulations with $\Lskin = 2000$ in Figure \ref{fig:bcurve_2e3}. The (anti)correlation between field strength and curvature has been well-studied in MHD turbulence (e.g., \citet{schek_structure, yang_curvature}); these studies find that $B \propto \Kpar^{-1/2}$ at high $\Kpar$, which corresponds to magnetic fields organized into structures that keep the magnetic tension force constant; we show this scaling with a red dotted line in Figure \ref{fig:bcurve_2e3}. However, in our PIC simulations, we find instead that $B \propto \Kpar^{-3/4}$ (blue dotted line) is a consistently better fit across each of our $\deltaB$. The high-$\Kpar$ tail steepens in the $\deltaB \approx 1$ and $2$ cases, showing significant deviation from both $\propto B^{-1/2}$ and $\propto B^{-3/4}$; this is better visualized in the projected curves of $\left<B(\Kpar)\right>/\Brms$ vs. $\Kpar$ shown in Figure \ref{fig:bcurve1D_2e3}. As we discuss in Section \ref{sec:aniso_curve}, the discrepancy between our measured $B$-$\Kpar$ correlation and the MHD expectation across all $\deltaB$ is likely due to the development of pressure anisotropy with respect to the magnetic field.

\begin{figure}[h]
    \includegraphics[width=0.5\textwidth]{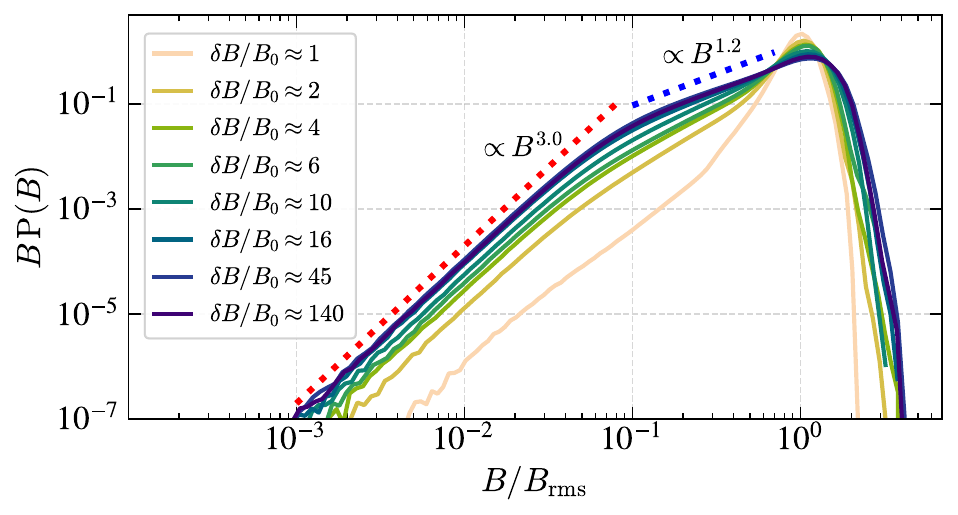}
    \caption{1D PDFs of $B/\Brms$, the local magnetic field strength normalized to the global rms magnetic field strength, for each of our simulations with $\Lskin = 2000$. The red dotted line indicates a scaling of ${\rm P}(B) \propto B^{2.0}$ (or $B \, {\rm P}(B) \propto B^{3.0}$) and the blue dotted line indicates ${\rm P}(B) \propto B^{0.2}$ (or $B \,{\rm P}(B) \propto B^{1.2}$).}
    \label{fig:bpdf_2e3}
\end{figure}

We can use our measurement of the $B$-$\Kpar$ correlation to derive an expected scaling for the distribution function of field-line curvature, ${\rm P}(\Kpar)$, at high $\Kpar$. We find that, at large $\deltaB$, the 1D PDF of magnetic field strength exhibits a ${\rm P}(B) \propto B^{1/5}$ scaling in the range $10^{-1} \lesssim B/\Brms \lesssim 1$ (see Figure \ref{fig:bpdf_2e3}), where the $B \propto \Kpar^{-3/4}$ correlation holds. We therefore have 
\begin{align} \label{eq:minustwo}
    &{\rm P}(\Kpar) \, {\rm d}\Kpar \propto {\rm P}(B) \, {\rm d}B \nonumber \\  &\implies {\rm P}(\Kpar) \propto \Kpar^{-19/10}.
\end{align}
This scaling is close to the theoretically-derived $\propto \Kpar^{-13/7}$ scaling expected in an MHD dynamo driven by a single-scale flow \citep{schek_dynamo}; however, this agreement may be a coincidence given that our simulations develop full turbulent cascades.  The slope of ${\rm P}(B)$ hardens slightly with decreasing $\deltaB$, reaching ${\rm P}(B) \propto B$ by $\deltaB \approx 4$; therefore, we expect a scaling closer to ${\rm P}(\Kpar) \propto \Kpar^{-2.5}$ for smaller $\deltaB$. This $\propto \Kpar^{-2.5}$ scaling has been observed both in MHD simulations \citep{yang_curvature} and in the solar wind \citep{riddhi, huang}. 

\begin{figure}[h]
    \includegraphics[width=0.5\textwidth]{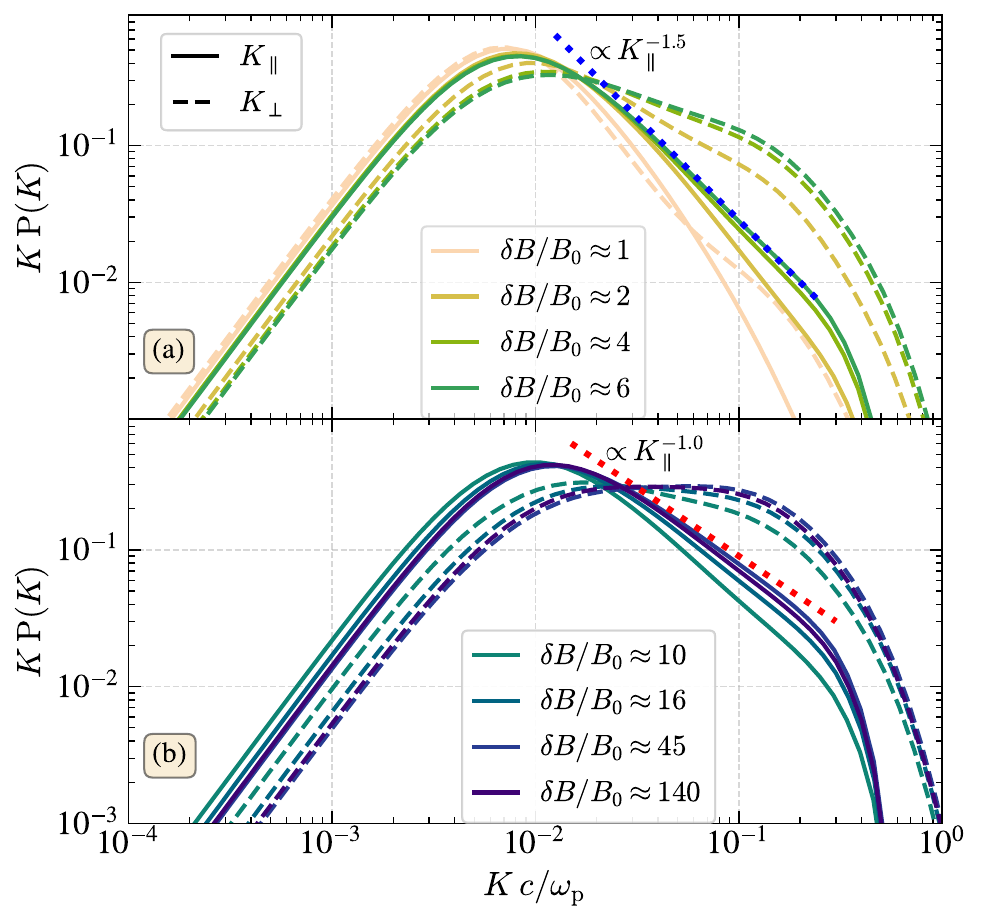}
    \caption{1D PDFs of $\Kpar$ (solid curves; Equation \eqref{eq:kpar}) and $\Kperp$ (dashed curves; Equation \eqref{eq:kperp}) for each of our simulations with $\Lskin = 2000$, split between the dynamo-like cases with $\deltaB \gtrsim 10$ (bottom panel) and the Kolmogorov-like cases with $\deltaB \lesssim 6$ (top panel). The red dotted line in the bottom panel indicates a power-law scaling  $\Kpar \, {\rm P}(\Kpar) \propto \Kpar^{-1}$ (or ${\rm P}(\Kpar) \propto \Kpar^{-2}$, as derived in Equation \eqref{eq:minustwo}) and the blue dotted line in the top panel indicates a scaling $\Kpar \, {\rm P}(\Kpar) \propto \Kpar^{-1.5}$ (or ${\rm P}(\Kpar) \propto \Kpar^{-2.5}$, as seen in MHD, e.g., \citet{yang_curvature}, and the solar wind, e.g., \citet{riddhi}).}
    \label{fig:curvepdf_2e3}
\end{figure}

In Figure \ref{fig:curvepdf_2e3}, we plot 1D PDFs of $\Kpar$ and $\Kperp$ for each of our simulations with $\Lskin = 2000$. For visualization purposes, we split the PDFs into two groups: the dynamo-like cases with $\deltaB \gtrsim 10$ (bottom panel) and the Kolmogorov-like cases (i.e., the cases affected by a dynamically important mean field, where the magnetic spectrum develops a Kolmogorov-like scaling) with $\deltaB \lesssim 6$ (top panel). Each of the PDFs of $\Kpar$ displays a prominent power-law tail extending towards high $\Kpar$, clearly illustrating the intermittency (i.e., non-Gaussianity) of sharp field-line bends; we show the ${\rm P}(\Kpar) \propto \Kpar^{-2}$ (or $\Kpar \,{\rm P}(\Kpar) \propto \Kpar^{-1}$) power-law scaling with a red dotted line and the ${\rm P}(\Kpar) \propto \Kpar^{-2.5}$ (or $\Kpar \,{\rm P}(\Kpar) \propto \Kpar^{-1.5}$) scaling with a blue dotted line. As expected, ${\rm P}(\Kpar) \propto \Kpar^{-2}$ is an excellent fit to our larger $\deltaB$ cases (especially $\deltaB \approx 140$), while ${\rm P}(\Kpar) \propto \Kpar^{-2.5}$ is an excellent fit to our smaller $\deltaB$ cases (especially $\deltaB \approx 6$). The $\Kpar$ PDFs for $\deltaB \approx 1$ and $2$ exhibit much steeper slopes than ${\rm P}(\Kpar) \propto \Kpar^{-2.5}$; these cases also stand out in our plots of $\left<B\right>$ vs. $\Kpar$ (Figure \ref{fig:bcurve1D_2e3}) and of ${\rm P}(B)$ (Figure \ref{fig:bpdf_2e3}), indicating that they are governed by different physics than the larger $\deltaB$ cases. 

In light of the above measurements, we caution that the power-law slope of the $\Kpar$ PDF depends on $\Lskin$; in Figure \ref{fig:curveres}, we show that, for smaller $\deltaB$, the power-law slopes for $\Lskin = 1000$ and $\Lskin=2000$ match nicely ($\deltaB \approx 1$ shows exceptional agreement between the power-law slopes across all three box sizes), while we require simulations at larger $\Lskin$ to confirm that our higher-$\deltaB$ cases are converged.  

The PDFs of $\Kperp$ strongly deviate from their MHD counterparts: the $\Kperp$ PDFs in MHD have rounded peaks that drop off smoothly towards the resistive scale \citep{kempski_1}, while our $\Kperp$ PDFs show shallow power-law slopes for $\deltaB \lesssim 10$ and remarkably flat, extended plateaus for $\deltaB \gtrsim 16$. These plateaus could be due to reconnection plasmoids with a broad size distribution; future work should determine the precise contribution of reconnection physics to the $\Kpar$ and $\Kperp$ PDFs. For each of the dynamo-like cases, the peak of the $\Kperp$ PDF lies at much smaller scales than that of the $\Kpar$ PDF, consistent with the MHD dynamo picture of elongated magnetic folds with large coherence lengths along the local magnetic field and small coherence lengths across the field \citep{schek_dynamo, rincon_dynamotheories, galish_tearing}; this contrasts with the small $\deltaB$ cases, for which the peaks of the $\Kpar$ and $\Kperp$ PDFs are nearly overlapping, suggesting that, for most bends, the coherence lengths along and across the field are similar.

% Previous MHD simulations show that the PDF of $\Kpar$ is sensitive to numerical grid resolution, with higher resolutions (i.e., smaller grid cells at a fixed box size) extending the power-law tail towards higher $\Kpar$, revealing increasingly tighter field-line bends; these simulations find no physical upper bound on $\Kpar$, even at resolutions up to $\left(10,240 \, {\rm cells}\right)^3$ \citep{kempski_1, kempski_2}. We perform a similar experiment in Figure \ref{fig:curveres}, where we keep the kinetic scale ($\skin$) fixed but vary the box size ($L$) between $\Lskin = 500$, $1000$, and $2000$, effectively varying the resolution. Our PDFs of $\Kpar$ are poorly converged for $\deltaB \gtrsim 10$: as we increase $\Lskin$, the power-law tails continually become steeper. However, for smaller $\deltaB$, the power-law slopes for $\Lskin = 1000$ and $\Lskin=2000$ match nicely; for $\deltaB \approx 1$, we find exceptional agreement between the power-law slopes across all three box sizes.     
% why steeper slope for larger box? does <K> shift predictably with resolution?

We can directly connect our statistics of $\Kpar$ to particle transport through collisionless large-amplitude turbulence. While we reserve a more comprehensive discussion of particle transport for Section \ref{sec:transport}, here we analyze statistics pertinent to a mode of particle scattering referred to as ``resonant curvature scattering''  \citep{kempski_1, lemoine_transport, lubke}: if a charged particle's gyroradius at a field-line bend is comparable to or slightly larger than the bend's radius of curvature, then the particle will see a full field reversal during a gyro-orbit, thus violating conservation of magnetic moment and yielding an order-unity change in the particle's pitch angle. Quantitatively, we write the condition for resonant curvature scattering as $\Kpar \, \gyro \gtrsim 1$, where
\begin{equation} \label{eq:gyro}
    \gyro \equiv \frac{\gamma mc^2}{qB}
\end{equation}
is the particle's instantaneous gyroradius in a magnetic field of strength $B$; $\gamma$ is the particle's Lorentz factor, $m$ is its mass, and $q$ is its charge. Since $\gyro$ at fixed $\gamma$ carries spatial-dependence through the local magnetic field strength, it is convenient to absorb
% the product $\Kpar \, \gyro$ cannot be meaningfully analyzed with the statistics of $\Kpar$ we have discussed above. However, we can absorb all 
this spatial dependence into the curvature statistics by defining the mean gyroradius,
\begin{equation} \label{eq:gyrobar}
    \overline{\gyro} \equiv \gyro \frac{B}{\Brms},
\end{equation}
and the renormalized curvature \citep{lemoine_transport}, 
\begin{equation} \label{eq:kparhat}
    \hat{K}_\parallel \equiv \Kpar \frac{\Brms}{B},
\end{equation}
such that the condition for resonant curvature scattering becomes
\begin{equation} \label{eq:curvescat}
    \Kpar \, \gyro = \Kparhatnoell \, \gyrobar \gtrsim 1.
\end{equation}
% Note that $\gyrobar$ depends only on $\gamma$ and thus -- in the absence of of particle acceleration -- is constant for a given particle energy.

To build PDFs that more accurately represent the probability of resonant curvature scattering, we further filter out curvature that does not contribute to scattering (e.g., field-line bends on scales much below the mean gyroradius or field-line kinks where high curvature only appears transiently). More precisely, we ``coarse grain'' the magnetic field below a scale $\ell$ (representing $\gyrobar$) by applying a uniform averaging filter of side-length $\ell$ to each cell in our box; we write the resulting smoothed and renormalized curvature as $\Kparhat$. In this way, integrating the 1D PDF of $\Kparhat \, \ell$ for $\Kparhat \, \ell \gtrsim 1$ yields a rough volume-filling fraction of viable scatterers for a particle with $\gyrobar \sim \ell$; we convert these volume-filling fractions to mean free paths in Section \ref{sec:transport}.

\begin{figure*}
    \includegraphics[width=1\textwidth]{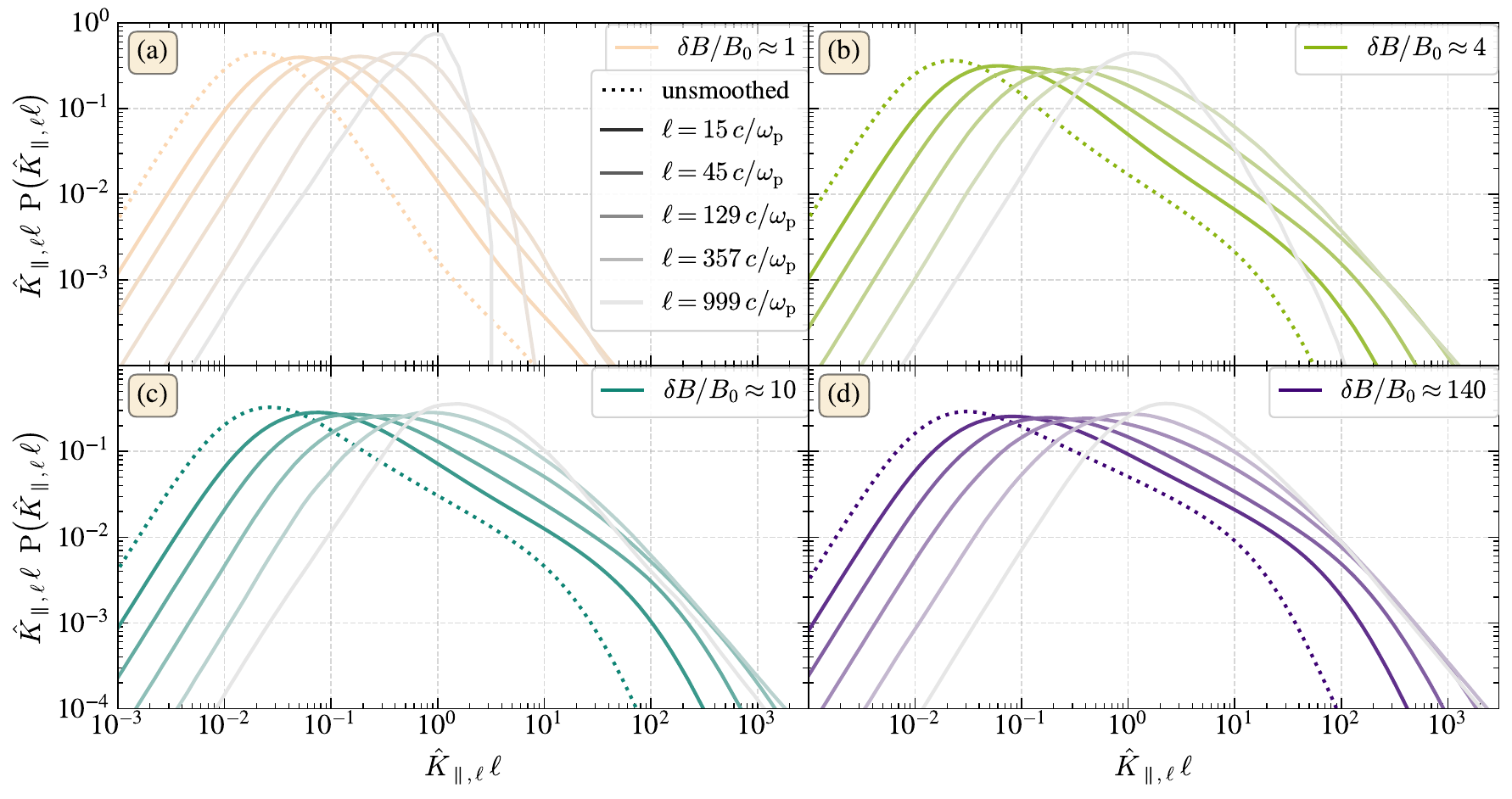}
    \caption{PDFs of the smoothed renormalized curvature, $\Kparhat$ (Equation \eqref{eq:kparhat}), multiplied by the smoothing scale, $\ell$, for our simulations with $\deltaB \approx 1$ (upper left), $4$ (upper right), $10$ (lower left), and $140$ (lower right) and $\Lskin=2000$. For each $\deltaB$, we show the PDFs of $\Kparhat \, \ell$ for five different smoothing scales, $\ell/(\skin) = 15$, $45$, $129$, $357$, and $999$, with lighter shades for larger $\ell$; the PDFs of renormalized curvature with no smoothing are depicted with dotted curves.}
    \label{fig:curvesmooth_2e3}
\end{figure*}

In Figure \ref{fig:curvesmooth_2e3}, we show 1D PDFs of $\Kparhat \, \ell$ for our $\deltaB \approx 1$ (upper left), $4$ (upper right), $10$ (bottom left), and $140$ (bottom right) simulations with $\Lskin = 2000$; for each $\deltaB$, we show the PDFs for five different smoothing scales, $\ell/(\skin) = 15$, $45$, $129$, $357$, and $999$. As we smooth at larger scales and remove more small-scale bends, the exponential cutoffs of the power-law tails of the PDFs gradually disappear; this smoothing produces nearly Gaussian PDFs at large $\ell$ for the small $\deltaB$ cases (i.e., $\deltaB \approx 1$ and $4$), yet it still leaves extended power-laws at larger $\deltaB$ (i.e., $\deltaB \approx 10$ and $140$). Additionally, the peak of these PDFs systematically shifts to lower $\Kparhat \, \ell$ as we increase $\ell$, as seen in MHD \citep{lemoine_transport}.
%We note that the peak of these PDFs also systematically shifts to lower $\Kparhat \, \ell$ as we increase $\ell$; these peaks scale tightly as $\left<\Kparhat \, \ell \right> \propto \ell^{-2/3}$, as also seen in MHD \citep{lemoine_transport}.
%If we were to show PDFs of just $\Kparhat$ (i.e., without the factor of $\ell$), we would also see that the peak of the PDF shifts towards smaller $\Kparhat$ as the smoothing scale increases.
% how does peak location scale with ell; compare to ?

In this section, we have found that certain properties of the intermittent magnetic field in collisionless large-amplitude turbulence -- such as the shape of the 1D PDF of field-line curvature, the $\Kpar^{-2.5}$ scaling of this PDF at high $\Kpar$ for moderate $\deltaB$, and the response of this PDF to coarse-graining -- broadly resemble their equivalent MHD measurements. However, these measurements generally disagree in detail, with small discrepancies in the correlation between $B$ and $\Kpar$ and, consequently, in the power-law slopes of the PDFs of $\Kpar$ for large $\deltaB$. In the next section, we explore what might be causing this disagreement.

\section{The impact of pressure anisotropy on collisionless large-amplitude turbulence} \label{sec:anisotropy}

A major difference between collisionless turbulence and its MHD counterpart is the ability of the former to generate pressure anisotropy. At $\beta \gtrsim 1$ -- which is the case for all of our simulations -- pressure anisotropy triggers kinetic instabilities \citep{instabilities_book} that can modify the properties of the turbulence. In Section \ref{sec:instabilities}, we provide evidence of these instabilities in our simulated turbulence; in Section \ref{sec:aniso_curve}, we show that, although these instabilities themselves do not significantly affect the statistics of $\Kpar$ and $\Kperp$, the underlying pressure anisotropy may explain the discrepancy with MHD detailed in Section \ref{sec:statistics}.

\subsection{Evidence of pressure-anisotropy-induced instabilities} \label{sec:instabilities}

In a collisionless magnetized plasma, changes in the magnetic field strength affect perpendicular and parallel particle motions in different ways due to the conservation of adiabatic invariants; macroscopically, this is quantified by the CGL equations \citep{cgl},
\begin{equation}
    \frac{P_{\perp}}{n B} = {\rm const.}
\end{equation}
and
\begin{equation}
    \frac{P_{\parallel} B^2}{n^3} = {\rm const.},
\end{equation}
where $P_{\perp}$ is the pressure perpendicular to the magnetic field, $P_{\parallel}$ is the pressure parallel to the field, and $n$ is the number density (for a single species). These equations tell us that, at fixed density, an increase in magnetic field strength causes an increase in $\Pperp$ but a decrease in $\Ppar$, generating pressure anisotropy. In the limit of incompressible turbulence, the magnetic field is strengthened when field lines are stretched and weakened when field lines are bent, so we expect positive pressure anisotropy (i.e., $\Pperp/\Ppar > 1$) in the former case and negative anisotropy (i.e., $\Pperp/\Ppar < 1$) in the latter.

Since our turbulence has $\beta \gtrsim 1$, we expect pressure anisotropy to trigger the mirror and firehose instabilities, which themselves regulate the anisotropy \citep{mirror, firehose, arzamasskiy}; mirror modes grow in regions of positive pressure anisotropy when
\begin{equation} \label{eq:mirror_thresh}
    \Pperp/\Ppar > 1 + 1/\betapar
\end{equation}
and (parallel) firehose modes grow in regions of negative pressure anisotropy when
\begin{equation} \label{eq:fire_thresh}
    \Pperp/\Ppar \lesssim 1 - 2/\betapar,
\end{equation}
where $\betapar \equiv \Ppar/(B^2/8\pi)$ is the ratio of parallel pressure to magnetic pressure. Mirror and firehose fluctuations grow fastest at scales comparable to the anisotropic species' thermal Larmor radius (in our case, the thermal electron or positron Larmor radius) and grow rapidly compared to the rate of macroscopic deformation; this allows the fluctuations to almost immediately feed back onto the flow \citep{squire1, squire2, majeski}.

\begin{figure*}
    \includegraphics[width=1\textwidth]{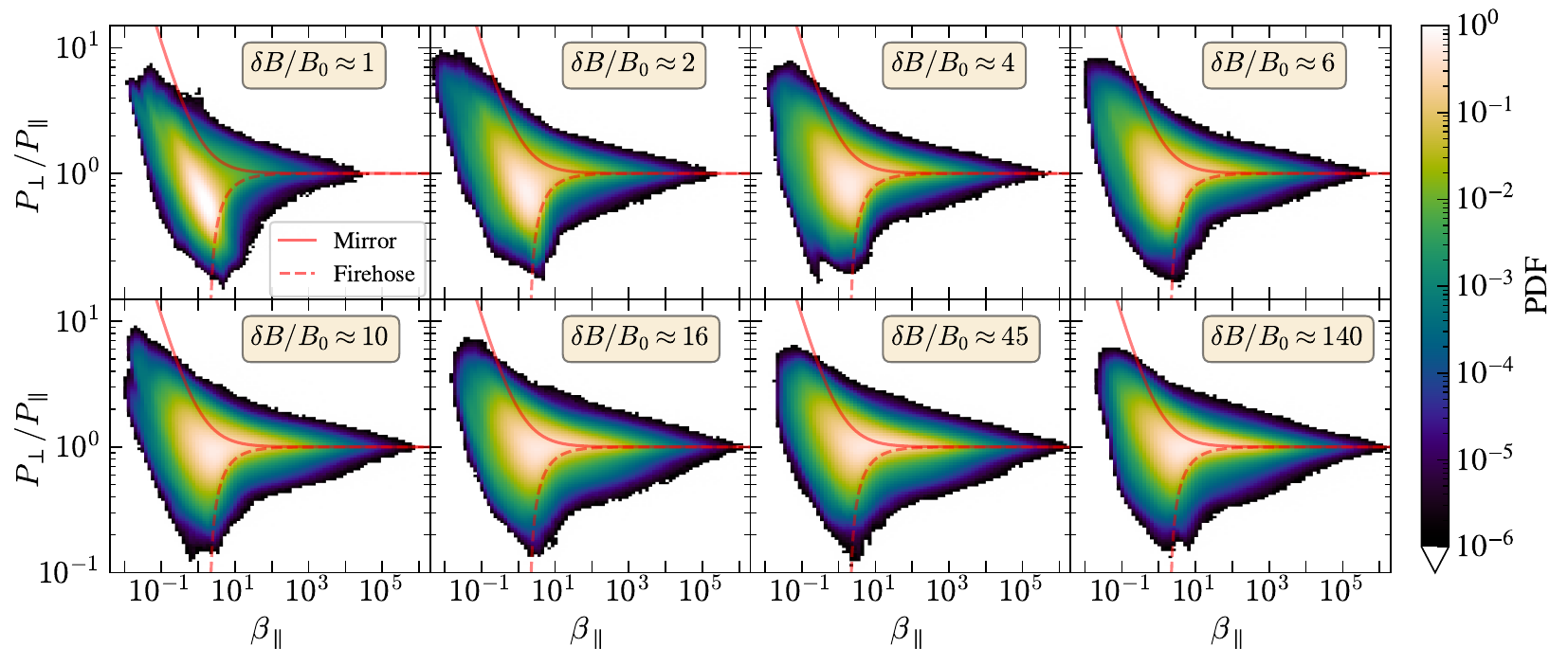}
    \caption{Brazil plots, or 2D PDFs of $\Pperp/\Ppar$ vs. $\betapar$, for each of our simulations with $\Lskin =2000$ (see panel labels). In each panel, the mirror instability threshold, $\Pperp/\Ppar > 1 + 1/\betapar$, is plotted as the solid red line and the firehose instability threshold, $\Pperp/\Ppar < 1 - 2/\betapar$, is plotted as the dashed red line.}
    \label{fig:brazilpdf_2e3}
\end{figure*}

One way in which mirror and firehose instabilities affect the plasma is by imposing a delimiting effect on the accessible phase space of the turbulence, pinning the system around their instability thresholds \citep{bale, rincon_dynamotheories}. We illustrate this effect in Figure \ref{fig:brazilpdf_2e3} by showing ``Brazil plots,'' or 2D PDFs of $\Pperp/\Ppar$ vs. $\betapar$, for each of our simulations with $\Lskin=2000$; we indicate the mirror threshold with a solid red curve and the firehose threshold with a dashed red curve. For each $\deltaB$, the vast majority of the probability density lies just within the stable region and a smaller but significant amount of the PDF straddles the instability thresholds. As $\deltaB$ decreases, the central contour shifts and distorts so that progressively less of the PDF lies outside the stable regions; by $\deltaB \approx 1$ (our lowest-$\beta$ case), we seem to enter a different regime where the central contour shows minimal correlation with the instability thresholds. 

% To further confirm the delimiting effects of mirror and firehose fluctuations, we have looked 

\begin{figure}[h]
    \includegraphics[width=0.5\textwidth]{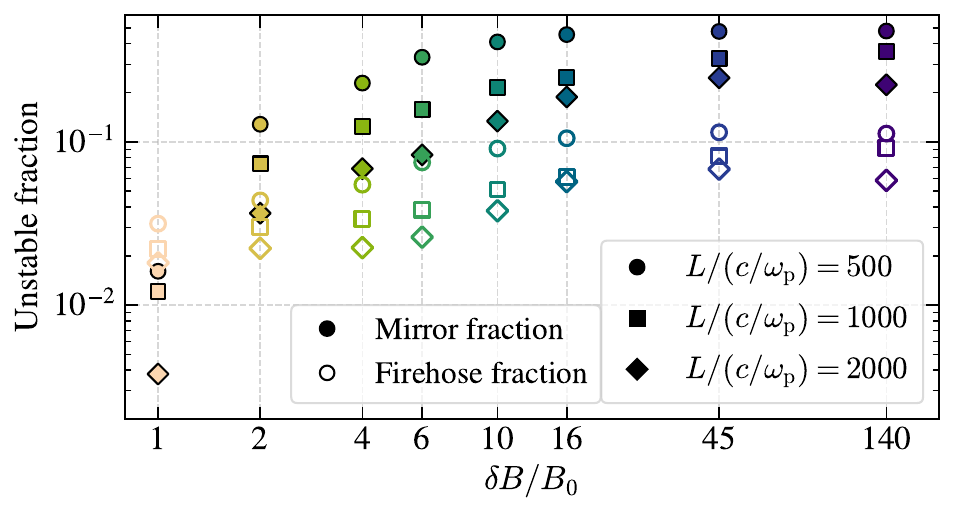}
    \caption{Volume-filling fractions of mirror-unstable (filled markers) and firehose-unstable (empty markers) regions in each of our simulations as a function of both $\deltaB$ (shown on the horizontal axis) and $\Lskin$ (differentiated by marker style); simulations with $\Lskin=500$ are represented by circular markers, those with $\Lskin=1000$ by square markers, and those with $\Lskin=2000$ by diamond markers.}
    \label{fig:instafrac}
\end{figure}

While Figure \ref{fig:brazilcurve_2e3} provides a qualitative picture of the volume-filling fractions of mirror- and firehose-unstable regions in our simulated turbulence, we quantify this in Figure \ref{fig:instafrac} as a function of both $\deltaB$ and $\Lskin$. For all $\deltaB$ except $\deltaB \approx 1$, the mirror-unstable fraction is larger than the firehose-unstable fraction by a factor of a few; in the large-$\deltaB$ dynamo-like cases with $\Lskin=2000$, $\sim 20\%$ of the volume is mirror-unstable and $\sim 5\%$ of the volume is firehose-unstable, with both of these fractions declining steadily for $\deltaB \lesssim 16$, giving $\sim4\%$ mirror-unstable and $\sim 2\%$ firehose-unstable for $\deltaB \approx 2$ and $\sim 0.4\%$ mirror-unstable and $\sim 2\%$ firehose-unstable for $\deltaB \approx 1$. Consistently, the unstable fraction decreases as $\Lskin$ increases; this could be because weaker driving (i.e., a smaller rate of strain, $\urms/L$) at larger $\Lskin$ is less effective at generating pressure anisotropy and can only push the plasma marginally past the instability thresholds. However, for $2 \lesssim \deltaB \lesssim 45$, there is evidence of asymptotic convergence with increasing $\Lskin$, since the difference in the unstable fraction is less pronounced between $\Lskin = 1000$ and $2000$ than it is between  $\Lskin = 500$ and $1000$.  

While mirror and firehose modes are responsible for regulating the pressure anisotropy, these are not the only possible Larmor-scale fluctuations triggered by the anisotropy. For example, cyclotron-type fluctuations like whistler waves -- which, like mirror modes, grow in regions of positive pressure anistropy \citep{sironi_cyclotron} -- are likely also present in our turbulence, but their impact is not obvious. Since these cyclotron-type fluctuations are usually expected for $\beta < 1$, they may be abundant in our turbulence with smaller $\deltaB$, possibly explaining why the central contour of the Brazil plot for $\deltaB \approx 1$ (in Figure \ref{fig:brazilpdf_2e3}) is not well-defined by the mirror and firehose thresholds. However, a quantitative investigation of the role of cyclotron-type instabilities in collisionless large-amplitude turbulence is outside the scope of this work and should be performed in an electron-ion plasma where ion cyclotron fluctuations and whistler fluctuations will not be degenerate in scale.

\subsection{Pressure anisotropy and field-line curvature} \label{sec:aniso_curve}

Growing mirror and firehose fluctuations each affect the morphology of the local magnetic field: where field lines are strengthened by the flow and become mirror unstable, perpendicular pressure forces cause field lines to bubble up, creating microscopic magnetic mirrors or bottles; where field lines are weakened by the flow and become firehose unstable, parallel pressure forces squeeze and pinch flux tubes, exacerbating tight field-line bends. Therefore, we expect these fluctuations to modify the field-line curvature throughout our simulated turbulence.

\begin{figure*}
    \includegraphics[width=1\textwidth]{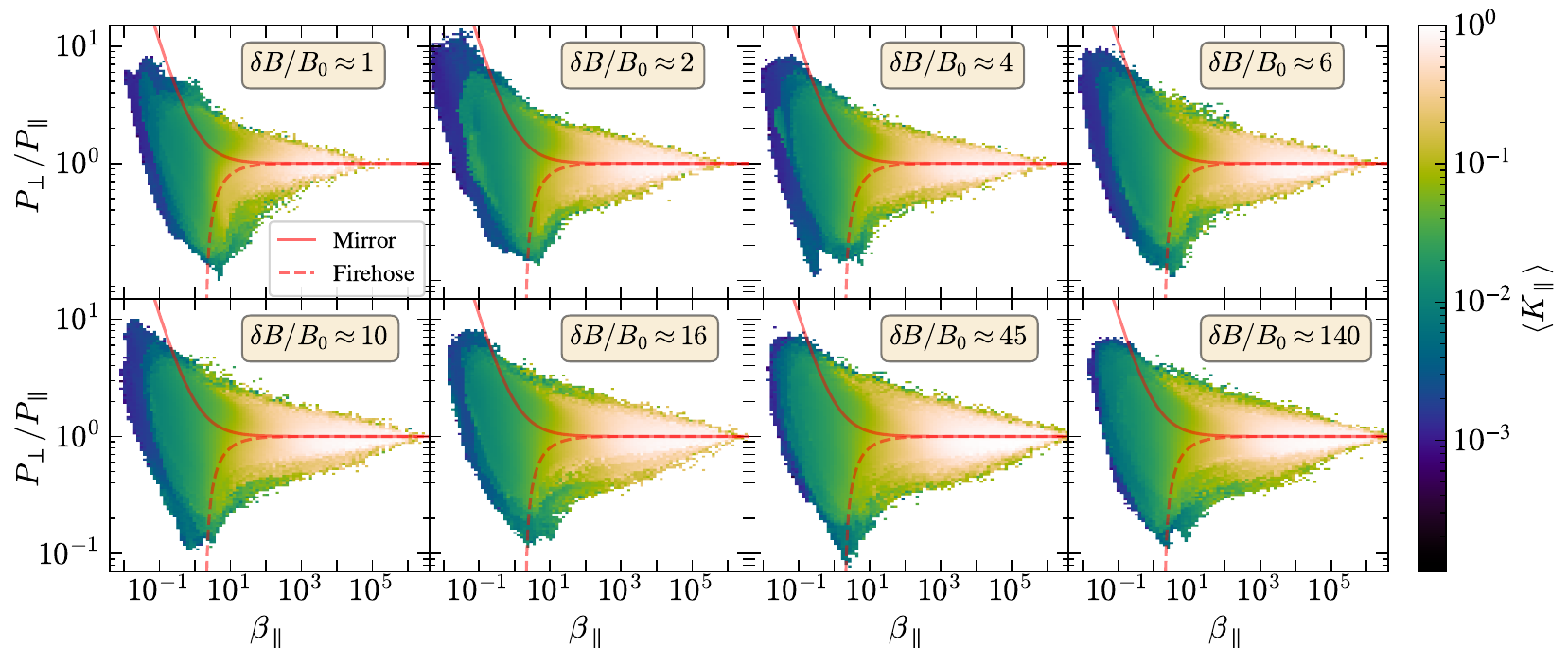}
    \caption{Brazil plots colored by local curvature, $\left<\Kpar\right>$, for each of our simulations with $\Lskin =2000$ (see panel labels). To maintain adequate dynamical contrast, the color range in each panel is normalized to the panel's maximum $\left<\Kpar\right>$. In each panel, the mirror instability threshold (Equation \eqref{eq:mirror_thresh}) is plotted as the solid red line and the firehose instability threshold (Equation \eqref{eq:fire_thresh}) is plotted as the dashed red line.}
    \label{fig:brazilcurve_2e3}
\end{figure*}

We can gain some qualitative insight into the correlation between pressure anisotropy and field-line curvature by making Brazil plots colored by the average $\Kpar$ in each bin; this is shown in Figure \ref{fig:brazilcurve_2e3} for each of our simulations with $\Lskin=2000$. Two patterns are recognizable across all $\deltaB$: first, curvature increases towards higher $\betapar$ (or, equivalently, towards lower magnetic field strength and weaker magnetic tension) where field lines are easier to bend; and second, curvature is enhanced in regions where the pressure anisotropy exceeds the mirror and firehose thresholds. These trends show little dependence on $\deltaB$, even down to $\deltaB \approx 1$. In other words, in each of our simulations, regions with intense curvature are co-spatial with mirror and firehose fluctuations. 

To gauge the impact of mirror and firehose fluctuations on the statistics of $\Kpar$, we perform a simple experiment in which we compute the 1D PDF of $\Kpar$ using only cells that are mirror- and firehose-\emph{stable}, but keep the overall sampling in $\betapar$ the same as in the full data set; in practice, we assign each cell a weight equal to ${\rm P}_{\rm total}(\betapar)/{\rm P}_{\rm stable}(\betapar)$ for that cell's local value of $\betapar$, where ${\rm P}_{\rm total}(\betapar)$ is the 1D PDF of $\betapar$ computed using the full domain and ${\rm P}_{\rm stable}(\betapar)$ is the PDF computed from only stable cells. We find that this reweighted PDF of $\Kpar$ in stable regions closely matches the global PDF of $\Kpar$, indicating that mirror and firehose-unstable regions do not contribute substantially to the PDF of $\Kpar$. This is likely due to the small volume-filling fractions of mirror and firehose fluctuations in our turbulence (Figure \ref{fig:instafrac}); we speculate that these fluctuations would play a more significant role at higher $\beta$, as seen in \citet{stonge_kunz}.

\begin{figure*}
    \includegraphics[width=1\textwidth]{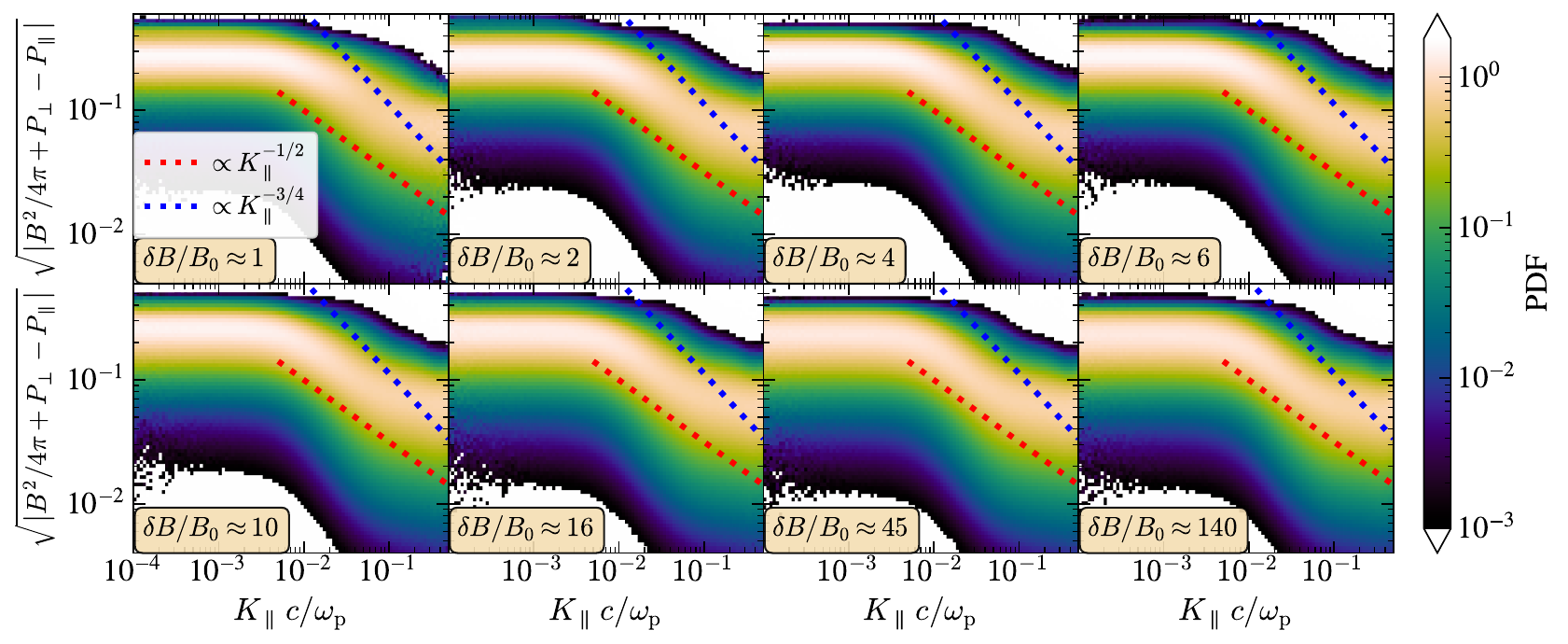}
    \caption{2D PDFs of the square root of the effective magnetic tension, $\sqrt{\left|B^2/4\pi + \Pperp - \Ppar \right|}$, vs. the local field-line curvature, $\Kpar$, for each of our simulations with $\Lskin=2000$ (see panel labels). The red dotted line in each panel shows the scaling $\propto \Kpar^{-1/2}$ while the blue dotted line shows $\propto \Kpar^{-3/4}$. To increase the contrast in the high-$\Kpar$ tails of these PDFs, we normalize the PDFs such that the integral over the 1D PDF of $\sqrt{\left|B^2/4\pi + \Pperp - \Ppar \right|}$ at each $\Kpar$ is unity.}
    \label{fig:tenscurve_2e3}
\end{figure*}

However, while the mirror and firehose fluctuations themselves do not significantly impact the statistics of curvature, the underlying pressure anisotropy can still have an effect. In a collisionless magnetized plasma, pressure anisotropy with respect to the magnetic field contributes to an effective field-line tension of the form $B^2/4\pi + \Pperp - \Ppar$, which can be derived by taking the first velocity moment of the Vlasov equation in the drift-kinetics limit \citep{kulsrud}. Since the $B \propto \Kpar^{-1/2}$ scaling in MHD is derived by balancing the magnetic tension with the velocity field \citep{schek_structure}, we expect that a similar scaling should hold in our collisionless turbulence if we incorporate the effective tension contributed by the pressure anisotropy. Therefore, instead of plotting 2D PDFs of $B$ vs. $\Kpar$ as we did in Figure \ref{fig:bcurve_2e3}, in Figure \ref{fig:tenscurve_2e3} we plot 2D PDFs of the square root of the effective field-line tension, $\sqrt{\left|B^2/4\pi + \Pperp - \Ppar \right|}$, vs. $\Kpar$. Across all $\deltaB$, the high-$\Kpar$ tail of the PDF is remarkably well-fit by the scaling $\sqrt{\left|B^2/4\pi + \Pperp - \Ppar \right|} \propto \Kpar^{-1/2}$, as expected. Evidently, pressure anisotropy accounts for at least some of the discrepancy in curvature statistics between collisionless large-amplitude turbulence and its MHD counterpart.

% We can perform a similar test for the scaling between the magnetic field strength and the curvature. In Figure \ref{fig:bcurveunstable_1min6_2e3}, we show 2D PDFs of $B/\Brms$ vs. $\Kpar$ computed over the total domain (top left), over just the mirror- and firehose-stable regions (top right), over just the mirror-unstable regions (bottom left), and over just the firehose-unstable regions (bottom right), for our simulation with $\deltaB \approx 140$ and $\Lskin=2000$. The slope of the central contour in the stable PDF closely matches the expected MHD scaling, $B \propto \Kpar^{-1/2}$ \citep{schek_dynamo}. However, the slope of the total PDF is pulled down to $B \propto \Kpar^{-2/3}$ because of the presence of mirror and firehose fluctuations in regions with weak magnetic fields and large curvature. Therefore, pressure-anisotropy-induced instabilities can readily explain the discrepancy between the MHD picture and the collisionless picture of large-amplitude turbulence.   

% \begin{figure*}
%     \includegraphics[width=1\textwidth]{fig_bcurveunstable_1min6_filter_2e3.pdf}
%     \caption{2D PDFs of the local magnetic field strength, $B/\Brms$, vs. the local curvature, $\Kpar$, computed over the total domain (top left), over just the mirror- and firehose-stable regions (top right), over just the mirror-unstable regions (bottom left), and over just the firehose-unstable regions (bottom right), for our simulation with $\deltaB \approx 140$ and $\Lskin=2000$. The dashed red line in the top left panel indicates the scaling $B \propto \Kpar^{-2/3}$, while the dotted red line in the top right panel indicates the scaling $B \propto \Kpar^{-1/2}$, as expected from MHD \citep{schek_dynamo}.}
%     \label{fig:bcurveunstable_1min6_2e3}
% \end{figure*}

\section{Discussion} \label{sec:discussion}

\subsection{Implications for particle transport in collisionless large-amplitude turbulence} \label{sec:transport}

% $\Kpar$ and $\Kperp$ are both directly relevant to particle transport through large-amplitude turbulence. For instance, if a particle encounters a field-line bend that curves on a scale comparable to the particle's local gyroradius, $r_g$ (i.e., if $K_\parallel r_g \sim 1$), the particle will resonantly scatter off the bend, experiencing an order-unity change in pitch angle. Or, if a particle encounters a field reversal on a scale comparable to its gyroradius (i.e., if $\Kperp r_g \sim 1$), the particle will drift away from the field line as its $\nabla B$-drift speed becomes comparable to its propagation speed. Therefore, it is useful to look at PDFs of $\Kpar$ and $\Kperp$ to quantify the range of curves or reversals that a particle might encounter.

Our results have immediate implications for particle transport through the turbulent collisionless plasma filling both the hot interstellar medium (ISM) of galaxies -- where $\deltaB$ can reach values of a few \citep{haverkorn} -- and the intracluster medium (ICM) of galaxy clusters -- where $\deltaB$ is likely large \citep{bonafede}. While we do not track individual particles or field lines in this work (though we plan to in follow-up work), the statistics we have collected allow us to make a few quantitative estimates pertinent to particle transport. Specifically, we can estimate the mean free path to non-adiabatic ``resonant'' curvature scattering and we can measure the probability of scattering off of microscale mirror fluctuations.
%we can quantify the correlation between curvature-dominated transport and large-scale magnetic mirroring, 

As a particle approaches a field-line bend, its fate is determined by its instantaneous gyroradius, $\gyro$ (Equation \eqref{eq:gyro}), and the curvature of the bend, $\Kpar$: if $\Kpar  r_g \gg 1$ then the particle will detach from the field line and drift away at the bend; if $\Kpar r_g \ll 1$ then the particle will stay attached to the field line and exactly follow the bend; and if $\Kpar r_g \gtrsim 1$ then the particle will sample a full field reversal in a single gyro-orbit, subsequently suffering an order-unity change in its pitch angle. Field-line curvature can therefore confine particles in two ways: by adiabatically guiding particles through field-line bends (when $\Kpar r_g \ll 1$) or by non-adiabatically (or ``resonantly'') scattering particles in pitch angle (when $\Kpar r_g \gtrsim 1$).

\begin{figure}
    \includegraphics[width=0.5\textwidth]{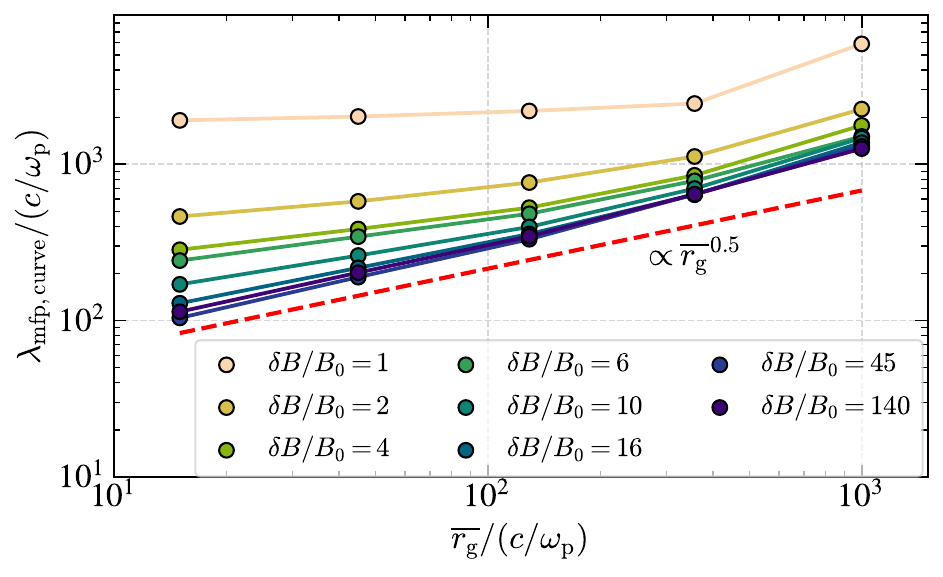}
    \caption{Mean free paths to resonant curvature scattering, $\lambdacurve$ (Equation \eqref{eq:mfp}), computed for $\ell/(\skin) \sim \gyrobar/(\skin) = 15$, $45$, $129$, $357$, and $999$ for each of our simulations with $\Lskin = 2000$. The dashed red line indicates a scaling $\lambdacurve \propto \gyrobar^{0.5}$, consistent with observations of Galactic cosmic rays.}
    \label{fig:mfps_2e3}
\end{figure}

As discussed towards the end of Section \ref{sec:statistics}, we can estimate a particle's mean free path to resonant curvature scattering by integrating over 1D PDFs of the renormalized curvature ($\Kparhatnoell$, Equation \eqref{eq:kparhat}) coarse-grained on a scale $\ell$; these PDFs of $\Kparhat \, \ell$ were shown in Figure \ref{fig:curvesmooth_2e3}. Taking $\Kparhat \, \ell \gtrsim 1$ to be the condition for resonant curvature scattering of a particle with $\gyrobar \sim \ell$ (for $\gyrobar$ the particle's mean gyroradius, Equation \eqref{eq:gyrobar}) -- meaning that the integral of the PDF of $\Kparhat \, \ell$ above unity gives the volume-filling fraction of regions where curvature scattering is possible -- we can estimate the mean free path to resonant curvature scattering with \citep{lemoine_transport}
\begin{equation} \label{eq:mfp}
    \lambdacurve = \frac{\ell}{\int_1^{\infty} {\rm P}_{\Kparhat \, \ell}\left(x \right) \, dx},
\end{equation}
where $x \equiv \Kparhat \, \ell$ and ${\rm P}_{\Kparhat \, \ell}\left(x \right)$ is the PDF of $x$. In Figure \ref{fig:mfps_2e3}, we show the mean free paths computed for $\ell/(\skin) \sim \gyrobar/(\skin) = 15$, $45$, $129$, $357$, and $999$ for each of our simulations with $\Lskin = 2000$. We find two salient trends in $\lambdacurve$. First, $\lambdacurve$ increases with decreasing $\deltaB$; this is consistent with the gradual reduction in probability density at high $\Kpar$ as $\deltaB$ decreases in Figure \ref{fig:curvepdf_2e3}. Second, $\lambdacurve$ increases with increasing $\gyrobar$, since coarse-graining at larger $\ell \sim \gyrobar$ removes more small-scale bends; this trend confirms the picture that particles with larger $\gyrobar$ can resonantly scatter off of fewer field-line bends. The positive trend between $\lambdacurve$ and $\gyrobar$ weakens with decreasing $\deltaB$, with $\deltaB \approx 1$ possibly showing a transition to a small-amplitude turbulence regime where curvature is more effective at scattering particles with larger $\gyrobar$. Given that $\gyrobar$ is proportional to a particle's energy, the trend of increasing $\lambdacurve$ with increasing $\gyrobar$ (for $\deltaB > 1$) is consistent with observations of cosmic rays in the Milky Way \citep{strong, aguilar_bc, amato}. In fact, our moderate- to large-$\deltaB$ cases exhibit trends close to $\lambdacurve \propto \gyrobar^{0.3 - 0.5}$; this is broadly consistent with observations of Galactic cosmic rays, although it requires that a significant fraction of the Galaxy is characterized by   $\deltaB \gtrsim $ a few. However, we caution that a standard mean free path may not be the most suitable metric for transport through magnetic field bends; recent high-resolution MHD simulations suggest that this transport may not be strictly diffusive, with particles exhibiting a wide distribution of scattering timescales \citep{kempski_2}.    

% While resonant curvature scattering is irrelevant for field-line bends on scales much larger than a particle's mean gyroradius, these large-scale structures can still affect particle transport through the random trajectories of field lines (which are followed by the particles) and through magnetic mirroring. While a full description of magnetic mirroring in collisionless large-amplitude turbulence is beyond the scope of this work, we show in Figure \ref{fig:mirror_2e3} that the field-line curvature and the mirror force,
% \begin{equation}
%     m \equiv B^{-1} \, \mathbf{\hat{b}} \cdot \left(\mathbf{\hat{b}} \cdot \nabla \right) \mathbf{B},
% \end{equation}
% are correlated. 

In collisionless plasmas with high $\beta$ (e.g., the ICM), thermal-Larmor-scale fluctuations are believed to also play a role in scattering and confining particles. In \citet{reichherzer}, the authors show that fluctuations seeded by the mirror instability (i.e. ``micromirrors'') are the dominant source of cosmic ray scattering and confinement up to $\sim {\rm TeV}$ energies; beyond this transition energy, cosmic rays mainly interact with structures in the inertial range of the ambient turbulence. While the TeV transition energy was computed assuming a $100\%$ volume-filling fraction for the micromirrors, the authors note that this energy should be reduced by a factor of $f_{\rm mm}^{1/(2-\delta)}$ for a more realistic filling fraction, $f_{\rm mm}$, and for $\delta$ set by the specific mesoscale turbulence model (e.g., $\delta=1/3$ corresponds to an isotropic Kolmogorov-like turbulence \citep{kolmogorov}, while $\delta=1/2$ corresponds to Iroshnikov-Kraichnan turbulence \citep{iroshnikov, kraichnan}). Taking $f_{\rm mm}$ to be $20\%$ (based on our $\deltaB \approx 140$ case with $\Lskin=2000$), the TeV transition energy is only reduced by a factor of $\sim 3$; for $f_{\rm mm} \approx 4\%$ (based on our $\deltaB \approx 2$ case with $\Lskin=2000$), the transition energy is reduced by a factor of roughly 10. We note that our measurements of the mirror-filling fraction are upper bounds, though we do not expect them to decrease much at higher $\Lskin$ given that they appear to be converging with $\Lskin$ in Figure \ref{fig:instafrac}; therefore, the TeV transition energy in the ICM appears robust assuming that the scale and amplitude of the micromirrors is as estimated in \citet{reichherzer}.

\subsection{Implications for particle acceleration in collisionless large-amplitude turbulence} \label{sec:acceleration}

The intermittent structures studied in this work -- namely, sharp magnetic field-line bends and 
%intense Larmor-scale magnetic fluctuations
rapid field reversals -- may allow collisionless large-amplitude turbulence to efficiently accelerate charged particles. In particular, we identify at least two mechanisms by which large-amplitude turbulence could energize particles: curvature-drift acceleration and magnetic pumping. 

The broad power-law distribution in field-line curvature present in our simulations has a significant impact on 
curvature-drift acceleration, whereby particles following a field line can gain or lose energy while moving through regions of changing magnetic field direction in the presence of a motional electric field \citep{drake, dahlin2014, dahlin2017, lemoine_21, bresci_22, lemoine_22, lemoine_25}. Recent work shows that curvature-drift acceleration is dominant over other drift-induced energization mechanisms, such as $\nabla B$-drift and polarization drift, as well as betatron acceleration in magnetically-dominated turbulence  \citep{sebastian}. Following that work, we can use as a proxy for the efficacy of curvature-drift acceleration the field-line contraction, 
\begin{equation} \label{eq:contract}
    \contract \equiv \mathbf{K} \cdot \mathbf{v_{\rm E}},
\end{equation}
where $\mathbf{K} \equiv \mathbf{\hat{b}} \cdot \nabla \mathbf{\hat{b}}$ is the vector curvature and  $\mathbf{v_{\rm E}} \equiv c(\mathbf{E} \times \mathbf{B}) / B^2$ is the $\mathbf{E} \times \mathbf{B}$ drift velocity (for $\mathbf{E}$ the electric field). If $\contract > 0$, the particle will see a contracting field line as it approaches a bend and energy from the released magnetic tension will be transferred to the particle (i.e., the particle will experience a ``head-on'' collision); if $\contract < 0$, the particle will see a stretching field line as it approaches a bend and energy will be transferred from the particle to the strengthening magnetic field (i.e., the particle will experience a ``tail-on'' collision). 

\begin{figure}
    \centering
    \includegraphics[width=0.5\textwidth]{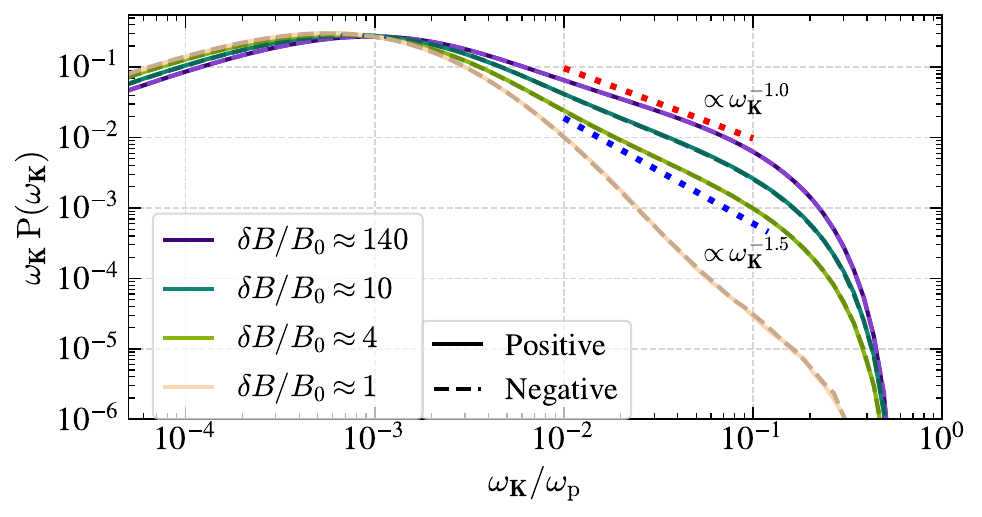}
    \caption{1D PDFs of field-line contraction, $\contract$ (Equation \eqref{eq:contract}) for our simulations with $\deltaB \approx 1$, $4$, $10$, and $140$ and $\Lskin=2000$; the PDFs of positive $\contract$ values are shown with solid lines while the PDFs of the magnitude of negative values are shown with dashed lines. The red dotted line indicates a scaling ${\rm P}(\contract) \propto \contract^{-2.0}$ (or $\contract \, {\rm P}(\contract) \propto \contract^{-1.0}$) while the blue dotted line indicates a scaling ${\rm P}(\contract) \propto \contract^{-2.5}$ (or $\contract {\rm P}(\contract) \propto \contract^{-1.5}$).}
    \label{fig:contraction}
\end{figure}

In Figure \ref{fig:contraction}, we show the 1D PDFs of field-line contraction, $\contract$, for our simulations with $\deltaB \approx 1$, $4$, $10$  and $140$; since we compute these PDFs in log space, we plot separate PDFs for positive $\contract$ (solid lines) and for the absolute value of negative $\contract$ (dashed lines). Each PDF possesses a power-law tail at large $\contract$, largely following the power-law tail at high field-line curvature; as in our 1D PDFs of $\Kpar$, the power-law becomes steeper as $\deltaB$ decreases, with the $\deltaB \approx 140$ PDF exhibiting a slope close to ${\rm P}(\contract) \propto {\rm P}(\Kpar) \propto \contract^{-2.0}$ and the $\deltaB \approx 4$ PDF exhibiting a slope close to ${\rm P}(\contract) \propto {\rm P}(\Kpar) \propto \contract^{-2.5}$. Additionally, the PDFs of positive and negative $\contract$ practically overlap, indicating that regions of field-line relaxation and field-line stretching are equally abundant in our turbulence; regardless, net particle energization can still result from stochastic energy exchange (i.e., second-order Fermi acceleration) despite there being no persistent net contraction. Given that the acceleration rate of curvature-drift acceleration follows $\dot{\gamma}/\gamma \propto \contract$, one should expect a distribution of acceleration rates that is strongly affected by intermittency, with the power-law tails in ${\rm P}(\contract)$ 
indicating more efficient particle acceleration at larger $\deltaB$.
Evidently, curvature-drift acceleration can be highly effective in large-amplitude turbulence. % mean acceleration rate is different; summary plot rms of distribution; % eventually, coarse-grained

In addition to curvature-drift acceleration, scattering off thermal-Larmor-scale fluctuations, such as those seeded by the mirror and firehose instabilities, can also heat and accelerate particles. In one such channel for particle energization called magnetic pumping, particle scattering off microscale fluctuations works in concert with the turbulent flow to energize particles \citep{berger1958,lichko2017,ley}: if the rate of scattering is comparable to the rate of bulk deformation of the plasma, an excess of energy can be transferred from particle motions perpendicular to the local magnetic field to motions parallel to the field due to the uneven response of the parallel and perpendicular pressures to adiabatic changes in the magnetic field. In our simulated electron-positron plasmas, thermal electrons and positrons can only be heated by the Larmor-scale fluctuations they seed; however, in an electron-ion plasma with sufficient scale separation between the electron and ion Larmor radii, nonthermal electrons can be accelerated by scattering off of ion-scale waves \citep{tran}.   

One particularly promising application of particle acceleration in collisionless large-amplitude turbulence is that of fossil electron re-acceleration in galaxy clusters. Some clusters, particularly disturbed or merging clusters, emit prominent MHz-GHz radio signals due to synchrotron and inverse-Compton cooling of energetic (1-100 GeV) cosmic ray electrons (CRe) over Myr-Gyr timescales \citep{van-weeren2019}. It is widely believed that these radiating CRe were re-accelerated from a population of ``fossil'' electrons, lower-energy (1-100 MeV) nonthermal electrons that do not emit detectable light and can remain dormant in the ICM for gigayears due to negligible collisional losses and minimal radiative losses at MeV energies \citep{ensslin1999,petrosian2001,pinzke2013}. The mechanism by which these fossil CRe are energized is still unclear, but the turbulence in which these CRe reside is almost certainly large-amplitude, since observations of energy equipartition in the ICM provide strong evidence of small-scale dynamo amplification \citep{bonafede,hitomi}. Future work should quantify the relative roles of curvature-drift, magnetic pumping, and other mechanisms in accelerating fossil electrons in the ICM.

\section{Summary and Conclusions} \label{sec:conclusions}

We performed fully kinetic 3D simulations of moderately subsonic, large-amplitude turbulence in a collisionless electron-positron plasma. By varying the strength of an initially weak mean magnetic field and allowing the plasma to reach a steady state under the influence of a stochastic driver, we were able to simulate steady-state turbulence with a wide range of $\deltaB$, from $\deltaB \approx 1$ up to $\deltaB \approx 140$ (which resembles the initially unmagnetized $\deltaB \rightarrow \infty$ case, analyzed in detail in \citet{sironi_dynamo}); this scan in mean-field strength also yielded a range of plasma $\beta$ in steady state, from $\beta \approx 1$ for $\deltaB \approx 1$ to $\beta \approx 4$ for $\deltaB \approx 140$. We also varied the scale separation between the turbulent driving scale and the kinetic scale, $\Lskin$, between 500, 1000, and 2000. In steady state, we analyzed the statistical properties of the intermittency in our simulated turbulence, focusing specifically on two classes of structures: sharp magnetic field-line bends (and rapid field reversals) and thermal-Larmor-scale magnetic field fluctuations (imprinted by the pressure-anisotropy-induced mirror and firehose instabilities). While we did not track individual particles or field lines in this work, our statistical analysis allowed us to infer the role of intermittent structures in both particle transport and acceleration. 

The main conclusions of our study are as follows:

\begin{itemize}
    %\item Intermittency is widespread in collisionless large-amplitude turbulence, appearing in each of our simulations from $\deltaB \approx 1$ to $140$.
    \item While MHD predicts a scaling between magnetic field strength ($B$) and scalar field-line curvature ($\Kpar$) of $B \propto \Kpar^{-1/2}$, the influence of pressure anisotropy in collisionless large-amplitude turbulence yields a steeper scaling, $B \propto \Kpar^{-3/4}$.
    \item The 1D PDF of $\Kpar$ displays an extended power-law tail for each $\deltaB$ (as also seen in MHD), indicating the abundant presence of intermittent field-line bends; this tail hardens with increasing $\deltaB$, exhibiting a slope $\propto \Kpar^{-2.5}$ (as also seen in MHD and the solar wind) for $\deltaB \approx 4$ and a slope $\propto \Kpar^{-2.0}$  for $\deltaB \approx 140$.  
    \item The shape of the 1D PDF of the perpendicular reversal scale, $\Kperp$, differs substantially from its MHD counterpart, displaying a power-law slope for smaller $\deltaB$ ($\deltaB \lesssim 10$) and an extended plateau for larger $\deltaB$ ($\deltaB \gtrsim 16$).
    \item Pressure-anisotropy-induced mirror and  firehose fluctuations are present in all of our simulations, but their volume-filling fractions decrease steadily towards smaller $\deltaB$; for $\deltaB \approx 140$, roughly $20\%$ of the volume is mirror-unstable and $6\%$ is firehose-unstable, while for $\deltaB \approx 2$, $4\%$ of the volume is mirror-unstable and $2\%$ is firehose-unstable. %While these mirror and firehose fluctuations are co-spatial with regions of high curvature, they do not significantly alter the curvature statistics.
    \item Both sharp field-line bends and Larmor-scale mirror fluctuations can affect the transport of charged particles through collisionless large-amplitude turbulence. By integrating over 1D PDFs of coarse-grained field-line curvature, we find that the particle mean free path to non-adiabatic (``resonant'') curvature scattering increases with particle energy (i.e., mean gyroradius) and decreases with $\deltaB$; for large $\deltaB$, this mean free path scales roughly with the square-root of energy. 
    \item Sharp field-line bends can accelerate well-magnetized particles via curvature-drift acceleration. The power-law tail of the distribution of field-line curvature yields a power-law tail in the distribution of acceleration rates due to curvature-drift acceleration, with the slopes of both distributions closely matching. 
\end{itemize}

Our investigation leaves a few avenues for future work. First, while pressure-anisotropy-induced instabilities did not significantly affect the curvature statistics in our present suite of simulations, one should investigate their impact in turbulence with higher $\beta$, where the volume-filling fractions of these fluctuations should be larger. Second, our simulations should be repeated in an electron-$ion$ plasma, allowing for the growth of ion-Larmor-scale instabilities like ion mirror, ion firehose, and ion cyclotron. Finally, instead of extracting information about particle transport and acceleration from statistical measurements, one should self-consistently track particle and field-line trajectories through kinetic simulations of collisionless large-amplitude turbulence.     

%% Please use the acknowkkledgment and contribution environments. This will 
%% be anonomyized when the "anonymous" style option is used. 
\begin{acknowledgments}
We thank Yan Yang, Riddhi Bandyopadhyay, and Muni Zhou for useful discussions related to this work. R.G. was supported by NASA FINESST grant 80NSSC24K1477. L.C. was supported by NSF PHY-2308944 and NASA ATP 80NSSC24K1230. L.S. was supported by DoE Early Career Award DE-SC0023015 and NSF grant PHY2409223. P.K. was supported by NASA ATP grant 80NSSC22K0667 and the Lyman Spitzer, Jr. Fellowship at Princeton University. This research was facilitated by Multimessenger Plasma Physics Center (MPPC) NSF grants PHY2206607 and PHY2206609. The work was also supported by a grant from the Simons Foundation to L.S. (MP-SCMPS-0000147).
Simulations were performed on NASA Pleiades (GID: s2356, s2610).
\end{acknowledgments}

% \begin{contribution}
% All authors contributed equally to this project.

%% Authors can use the Contributor Role Taxonomy (CRediT) at
%% for ideas on how write a good statement tailored to their needs.

% \end{contribution}

\software{TRISTAN-MP \citep{tristan}}

\appendix

\section{Dependence on scale separation}

In this section, we discuss the dependence on scale separation (or, equivalently, simulation box size), $\Lskin$, for several quantities analyzed in the main text.

\begin{figure}[h]
    \includegraphics[width=0.5\textwidth]{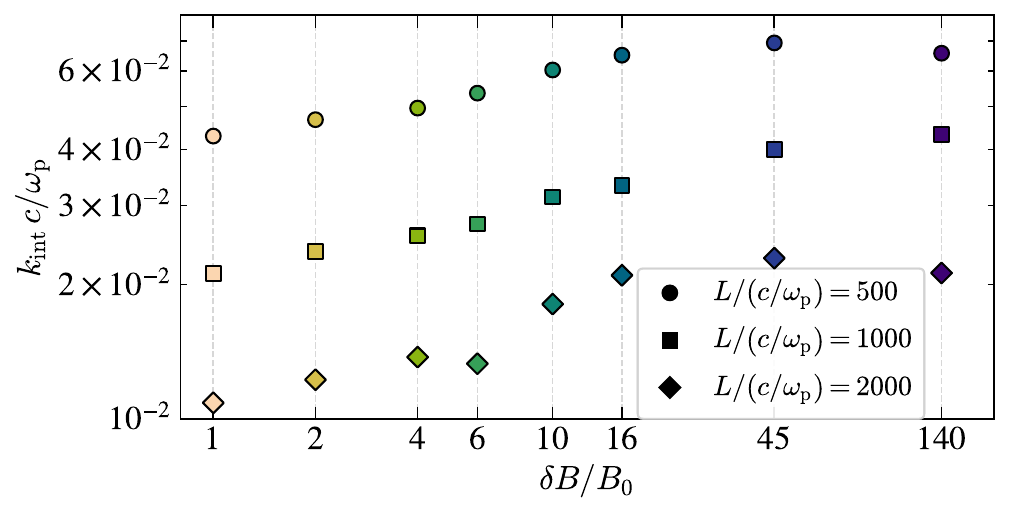}
    \caption{Integral-scale wavenumber (Equation \eqref{eq:kint}) as a function of both $\deltaB$ (horizontal axis) and $\Lskin$ (marker shape); circles correspond to $\Lskin = 500$, squares to $\Lskin=1000$, and diamonds to $\Lskin=2000$.}
    \label{fig:kints}
\end{figure}

Figure \ref{fig:kints} shows the integral-scale wavenumber of the turbulent magnetic field, $k_{\rm int}$ (Equation \eqref{eq:kint}), as a function of both $\deltaB$ (horizontal axis) and $\Lskin$ (marker shape). For each $\deltaB$, increasing $\Lskin$ by a factor of two decreases $k_{\rm int}$ by a factor of two (i.e., increases the dominant scale of the magnetic field by a factor of two), as expected. 

% Previous MHD simulations show that the PDF of $\Kpar$ is sensitive to numerical grid resolution, with higher resolutions (i.e., smaller grid cells at a fixed box size) extending the power-law tail towards higher $\Kpar$, revealing increasingly tighter field-line bends; these simulations find no physical upper bound on $\Kpar$, even at resolutions up to $\left(10,240 \, {\rm cells}\right)^3$ \citep{kempski_1, kempski_2}. We perform a similar experiment in Figure \ref{fig:curveres}, where we keep the kinetic scale ($\skin$) fixed but vary the box size ($L$) between $\Lskin = 500$, $1000$, and $2000$, effectively varying the resolution. 

\begin{figure*}
    \includegraphics[width=1\textwidth]{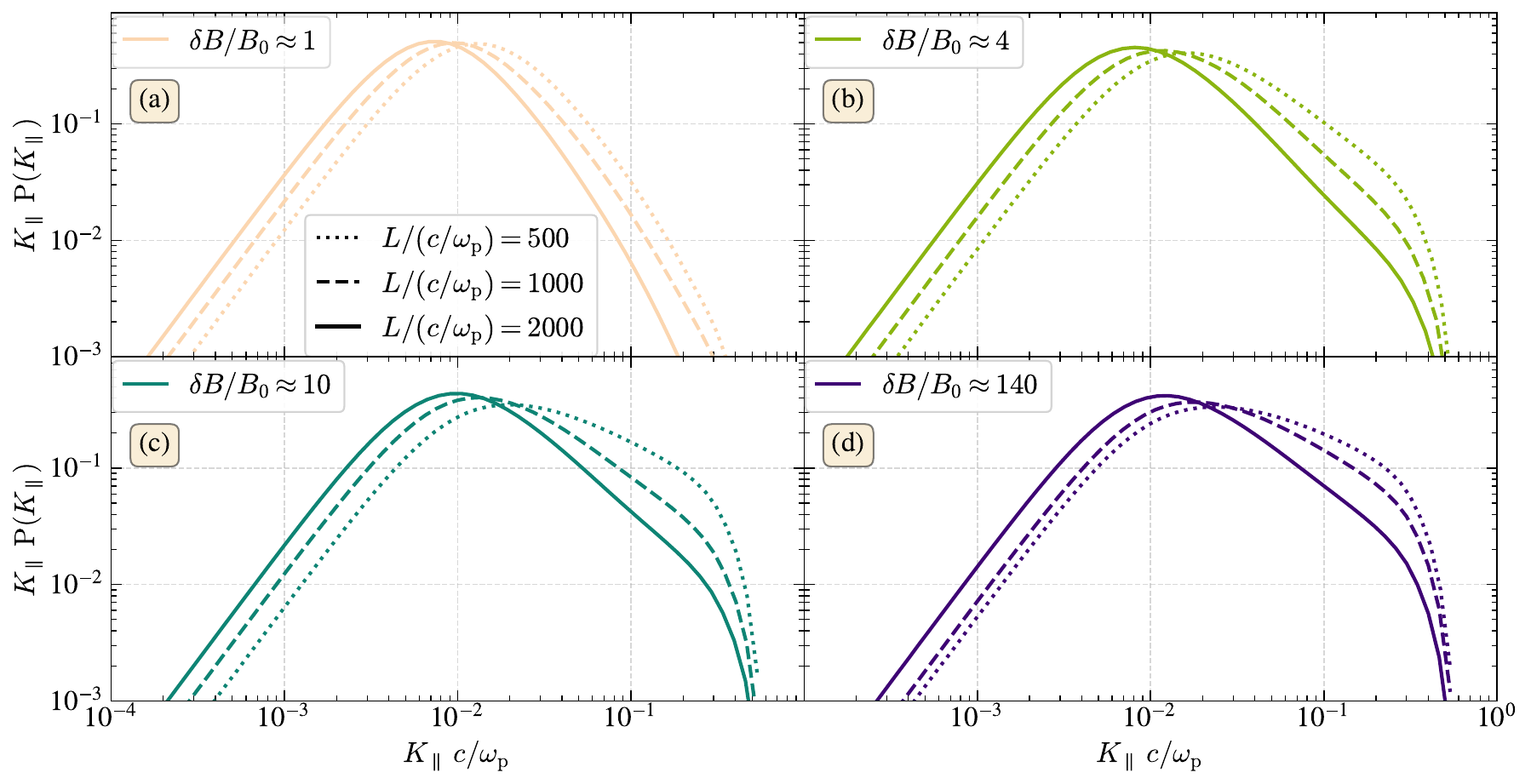}
    \caption{PDFs of $\Kpar$ as a function of $\Lskin$ for our simulations with $\deltaB \approx 1$ (top left), $4$ (top right), $10$ (bottom left), and $140$ (bottom right). In each panel, the dotted curve corresponds to $\Lskin = 500$, the dashed curve to $\Lskin = 1000$, and the solid curve to $\Lskin = 2000$.}
    \label{fig:curveres}
\end{figure*}

Figure \ref{fig:curveres} shows 1D PDFs of $\Kpar$ for our simulations with $\deltaB \approx 1$ (top left), $4$ (top right), $10$ (bottom left), and $140$ (bottom right); in each panel, the dotted curve corresponds to $\Lskin = 500$, the dashed curve to $\Lskin = 1000$, and the solid curve to $\Lskin = 2000$. For $\deltaB \gtrsim 10$, the power-law slopes in our PDFs of $\Kpar$ are not yet converged: as we increase $\Lskin$, the power-law tails continually become steeper. However, for smaller $\deltaB$, the power-law slopes for $\Lskin = 1000$ and $\Lskin=2000$ match nicely; for $\deltaB \approx 1$, we find exceptional agreement between the power-law slopes across all three box sizes.

\bibliography{main.bib}{}
\bibliographystyle{aasjournalv7}

\end{document}